\documentclass[12pt,a4paper,one-sided,openright]{report}
\usepackage{setspace}
\usepackage{graphicx}
\usepackage{amsmath}
\usepackage{fancyhdr}
\usepackage{geometry}

\newcommand{\apj}{ApJ}
\newcommand{\aap}{A\&AP}
\newcommand{\aj}{Astro. J.}
\newcommand{\apjl}{ApJ Lett.}

\newcommand{\apjs}{ApJ Suppl.}

\newcommand{\MNRAS}{MNRAS}
\newcommand{\mnras}{MNRAS}

\newcommand{\pra}{Phys. Rev. A}
\newcommand{\prd}{Phys. Rev. D}
\newcommand{\physrep}{Phys. Rep.}
\newcommand{\be}{\begin{equation}}
\newcommand{\ee}{\end{equation}}
\newcommand{\ba}{\begin{eqnarray}}
\newcommand{\ea}{\end{eqnarray}}
\newcommand{\brr}{\begin{array}}
\newcommand{\err}{\end{array}}
\newcommand{\bc}{\begin{center}}
\newcommand{\ec}{\end{center}}

\newcommand{\hm}{\,h^{-1}{\rm Mpc}}

\newcommand{\vvec}{{\bf v}}
\newcommand{\xvec}{{\bf x}}
\newcommand{\rvec}{{\bf r}}
\newcommand{\gvec}{{\bf g}}
\newcommand{\kvec}{{\bf k}}

\newcommand{\yvec}{{\bf y}}
\def\etal {\rm {\it et al.}}
\def\h0 {\rm H_{0} }

\def\kms{\ifmmode\,{\rm km}\,{\rm s}^{-1}\else km$\,$s$^{-1}$\fi}
\def\hmpc{\ifmmode\,{\it h }^{-1}\,{\rm Mpc }\else $h^{-1}\,$Mpc\,\fi}

\def\\{\hfill\break}

\def\pscz{{\it PSC}z}

\def\cf{{\it cf.}}
\def\eg{{\it e.g.}}
\def\ie{{\it i.e.}}
\def\kms{\ifmmode\,{\rm km}\,{\rm s}^{-1}\else km$\,$s$^{-1}$\fi} 
\def\kms{\,{\rm km\,s${^{-1}}$}}

\def\ltsima{$\; \buildrel < \over \sim \;$}
\def\lsim{\lower.5ex\hbox{\ltsima}}
\def\gtsima{$\; \buildrel > \over \sim \;$}
\def\gsim{\lower.5ex\hbox{\gtsima}}

 \def\v0{\pmb{$0$}}

\begin{document}

\pagenumbering{arabic}

\setcounter{page}{1}

\newpage

\thispagestyle{empty}
\begin{center}
  {\LARGE \bf Probing the Large-Scale Homogeneity of the Universe with Galaxy Redshift Surveys} \\
  \vspace*{2cm}
  {\Large  Cristiano Giovanni Sabiu, BSc}

  \vspace*{1cm}

  { Thesis \\ submitted to the \\ University of Glasgow \\ 
	                for the degree of \\ MSc}
  \vspace*{0.9cm}

   \begin{center}
   \includegraphics{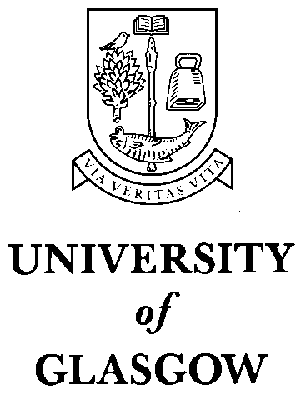}
   \end{center}

  \vspace*{0.2cm}

  {\large Astronomy \& Astrophysics Group\\
          [-1mm] Department of Physics \& Astronomy\\
          [-1mm] University of Glasgow\\
          [-1mm] Scotland\\
           [1mm] September 2006}

\end{center}
\vspace{1cm}
\begin{flushright}
{\copyright \  Cristiano Sabiu 2006}
\end{flushright}

\newpage
\thispagestyle{empty}
\begin{center}
\vspace*{5cm}
\textit{\LARGE {dedicato alla \\[2mm]
\qquad\qquad mia nonna \\[2mm]
\qquad\qquad\qquad \& \\[2mm]
\qquad\qquad\qquad\qquad\quad al mio nonno}}\\
 
\end{center}

\newpage
\addcontentsline{toc}{chapter}{\numberline{}Abstract}
\begin{center}
  \textbf{\Large Probing the Large-Scale Homogeneity of the Universe with Galaxy Redshift Surveys}

  \vspace*{1cm}
  \textbf{\large Cristiano Sabiu BSc}

  \vspace*{0.5cm}
  {\large Submitted for the degree of MSc\\ September 2006}

  \vspace*{1cm}
  \textbf{\large Abstract}
\end{center}
\begin{spacing}{1.5}
Modern cosmological observations clearly reveal that the universe contains a
hierarchy of clustering.  However, recent surveys show a transition to
homogeneity on large scales. The exact scale at which this transition occurs
is still a topic of much debate.  There has been much work done in trying to
characterise the galaxy distribution using multifractals.  However, for a
number of years the size, depth and accuracy of galaxy surveys was regarded
as insufficient to give a definitive answer.  One of the main problems which
arises in a multifractal analysis is how to deal with observational selection
effects: \ie~`masks' in the survey region and a geometric boundary to the
survey itself.

In this thesis I will introduce a {\em volume\/} boundary correction which is 
rather similar
to the approach developed by Pan and Coles in 2001, but which improves on
their {\em angular\/} boundary correction in two important respects: firstly,
our volume correction `throws away' fewer galaxies close the boundary of a 
given data set and secondly it is computationally more efficient.

After application of our volume correction, I will then show how the underlying 
generalised dimensions of a given point set can be computed.  I will apply this
procedure to calculate the generalised fractal dimensions of both simulated 
fractal point sets and mock galaxy surveys which mimic the properties of the 
recent IRAS PSCz catalogue.
\end{spacing}

\chapter*{Acknowledgments}
\addcontentsline{toc}{chapter}{\numberline{}Acknowledgments}

Firstly, I must thank my two supervisors Dr. Martin Hendry \& Dr. Luis Teodoro 
who have kept me on the right track throughout my research.  
\\

\noindent I express my gratitude to my parents, Sheena \& Antonio, and  my grand parents 
Maria-Elena \& Giovanni Sabiu, who have been a constant support to me through my 
whole life: This thesis is for you.
\\ 

\noindent To my friend (\& office-mate) Luis, thank you for teaching me something 
new each day. I hope we will work on interesting things together in the future.
\\

\noindent Thanks also to some great guys, Paul Holt \& Eamon Scullion; 
what can I say? Thanks for letting me holiday in the surreal world you two live in. 
Alejandro `Speedy' Gonzalez, Russell `Render' Johnston \& Robert McKay.
\\

\noindent Finally, in the spirit of leaving the best till last, I wish to thank my 
fianc\'{e}e (at the time of writing) Georgie Mooney for sharing her life with me. 

\chapter*{Declaration}
\addcontentsline{toc}{chapter}{\numberline{}Declaration} The work in
this thesis is based on research carried out at the Department of Physics 
\& Astronomy within the Astronomy \& Astrophysics Group of the University 
of Glasgow, Scotland.  No part of this thesis has been
submitted elsewhere for any other degree or qualification and it is all
my own work unless referenced to the contrary in the text.

\vspace{2in}
\noindent \textbf{Copyright \copyright  2006 by Cristiano Sabiu}.\\
``The copyright of this thesis rests with the author.  No quotations
from it should be published without the author's prior written
consent and information derived from it should be acknowledged.''

 \tableofcontents
 \listoffigures

\chapter*{Abbreviations}
\addcontentsline{toc}{chapter}{\numberline{}Abbreviations}

\begin{tabular}{ll}
\hline \hline \\[2mm]
2dF & 2 degree field (galaxy redshift survey) \\
2PCF & Two-Point Correlation Function \\
BAO & Baryon Acoustic Oscillations \\
CDM & Cold Dark Matter \\
CfA & Center for Astrophysics \\
CMB & Cosmic Microwave Background \\
CP & Cosmological Principle \\
Gpc & Gigaparsec \\
ISM & Interstellar Medium \\
LG & Local Group \\
LSS & Large Scale Structure \\
Mpc & Megaparsec \\
MST & Minimal Spanning Trees \\
\pscz & Point Source Catalogue (redshift) \\
RWM & Robertson Walker Metric \\
SDSS & Sloan Digital Sky Survey \\
WMAP & Wilkinson Microwave Anisotropy Probe \\[5mm]
\hline \hline
\end{tabular}

\clearpage

\pagestyle{fancy}
\renewcommand{\sectionmark}[1]{\markright{\thesection\ #1}}
\fancyhf{}
\fancyfoot[C]{\bfseries\thepage}
\fancyhead[LE]{\em \leftmark}
\fancyhead[RO]{\em \leftmark}
\renewcommand{\headrulewidth}{0.5pt}
\renewcommand{\footrulewidth}{0pt}
\addtolength{\headheight}{0.5pt}
\fancypagestyle{plain}{%
        \fancyhead{}
        \renewcommand{\headrulewidth}{0pt}
}
\parskip=12pt
\topmargin=0.0in
\textheight = 8.7in
\textwidth = 170mm
\marginparsep=0pt
\marginparwidth = 5mm
\oddsidemargin = 13mm
\footskip = 40pt 

\begin{spacing}{1.5}
\chapter{Introduction}
\vspace{-0.6cm}
This thesis is devoted to a study of the large scale structure (LSS) 
of the universe and as such belongs in the field of Cosmology. Cosmology is the study 
of structure and evolution in the universe. The main constituents of LSS are individual 
galaxies and clusters of galaxies up to Gigaparsec (Gpc) scales. 
\section{Cosmology through the ages}
The ancient Greeks were undoubtedly the leaders in astronomical understanding 
of their time. Around the 4th century BC a general consensus emerged, from the 
combined ideas of many philosophers, including Plato and Aristotle, which put our 
spherical Earth at the centre of the universe. They speculated that the Sun, Moon and 
planets were carried around the Earth on concentric spheres, arranged: Moon, Sun, Venus, Mercury, 
Mars, Jupiter, Saturn and the fixed stars beyond. Aristotle was to later elaborate 
on this {\em geocentric} model by trying to explain the lunar cycle. 

An updated {\em geocentric} theory of the heavens was put together by the astronomer 
Claudius Ptolemy from many works in Greek astronomy. His {\em Ptolemaic} model was 
penned in the 2nd century AD and stood as the standard theory for more than a 
millennium. Ptolemy made extensive use of epicycles to explain many aspects of 
planetary motion. In particular, his epicyclic explanation of retrograde motion in the 
planets helped elevate this theory to the forefront of astronomical thinking. 
This picture of our universe stood solid until the 16th century when Nicolaus 
Copernicus changed forever our view of the cosmos. 

Copernicus, the famous Polish astronomer, showed that a model, with the sun at its centre, could explain 
the motion of the planets in a very simple way, with no need for complicated 
orbits and epicycles. However this {\em Heliocentric} model was not new at all: 
its origins dated back many centuries BC to the workings of an Indian 
philosopher, Yajnavalkya. He had the vision to see that the sun, being the 
most important of the heavenly bodies, should be at the centre of our universe. 
However he lacked any real observational or scientific evidence. 
This Heliocentric idea was also present in ancient Greece, held strong by the 
Pythagoreans. The first to propose this was Aristarchus of Samos (c. 270 BC) 
and later Archimedes, the Greek scientist, was swayed by this line of thinking. 

\section{The Cosmological Principle}\label{CP}
Despite the greater simplicity involved with the Heliocentric system, it would 
not come to dominate the astronomical community. This was, at least in part, not for scientific 
reason but religious prejudices. Many religions held the false belief that the Earth 
was somehow special and therefore must be the centre of our observable universe. 
The Roman Catholic Church had a strong hold on science and particularly astronomy 
(due to its close connection to the heavens). Any theories which did not conform 
to the teachings of the Bible were deemed `heretical' and were hidden away 
from public knowledge. Despite this dogma, Copernicus in the 16th century 
managed to garner much support for the Heliocentric model, mainly due to his 
scientific writings, {\em De Revolutionibus} (1543) and Galileo's supporting 
observations. Galileo later opposed the Catholic Church by his strong support for the 
{\em Copernican} ideas. While on trial for heresy he famously said of the Earth, 
``Eppur si muove'' - and yet it moves\footnote{Galileo probably never spoke these 
exact words, however they stand as a symbol of his support for scientific truth.}.  

So why are we dwelling in the past here? The reason is this; Copernicus and Galileo 
did not see themselves at the centre of the solar system. They had displaced everyone 
from a special location in space and this trend would not stop with the Earth. It would 
later lead to the whole solar system being placed in the outer rim of our Milky way 
galaxy. Then, in the early 20$^{th}$ century our galaxy became one of many. Now we are 
but a mere speck of dust in a vast, ever expanding universe, a fact that was first 
realised by the American astronomer Edwin Hubble in the 1920's.

\subsection*{Edwin Hubble}
Edwin Hubble studied the systematic variations of the red-shift in the `spiral 
nebulae', as they were known. The redshift $z$ is defined as,
\be
z=\frac{\lambda_{O}}{\lambda_{E}}-1,
\ee
where $\lambda_{O}$ and $\lambda_{E}$ are the observed and emitted wavelengths of light, 
respectively. Hubble used this technique to investigate populations of similar objects, 
usually galaxies of a particular morphological type, and examined the relationship between 
the red-shifts and their relative brightnesses. What he found is now known as 
Hubble's Law: the red-shift in the spectra of the objects grow as the objects 
became more distant. The farther away an object, the faster it is receding from 
us. This, Hubble concluded, is because the universe itself is expanding -- a fact that 
(as we will see in the next Chapter) was consistent with the theoretical predictions of 
Einstein's General Theory of Relativity

In fact Hubble found this connection to be a linear relation between recessional 
velocity and distance. The usual Hubble law is written as,
\be
v = H_{0}D,
\ee
where $v$ is the recessional velocity due to redshift, typically expressed in km/s. 
$H_{0}$ is Hubble's constant and corresponds to the value of $H$ (often termed the 
Hubble parameter which is a value that is time dependent) in the Friedmann equations 
(\cf~eq.\ref{eq:fried1}) taken at the time of observation denoted by the subscript 0. 
This value is the same throughout the universe for a given conformal time. $D$ is the 
proper distance that the light had travelled from the galaxy in the rest frame of the 
observer, measured in megaparsecs (Mpc).

For relatively nearby galaxies (\ie~$z\ll{1}$), the velocity $v$ can be estimated from the galaxy's 
redshift $z$ using the formula $v = zc$ where $c$ is the speed of light. For more distant galaxies, 
the relationship between recession velocity and distance becomes more complicated and requires general 
relativity (see Ch.2).

In using Hubble's law to determine distances, only the velocity due to the expansion of the 
universe should be used. Since gravitationally interacting galaxies move relative to each other 
independent of the expansion of the universe, these relative velocities, called peculiar 
velocities, need to be accounted when applying Hubble's law. So more generally the Hubble 
law is,
\be
v_{rec}={\rm H_{0}}D+v_{pec}.
\ee
In this case, $v_{pec}$ is the radial component of the peculiar motion of the object. 
As an example the local group of galaxies has a $v_{pec}\approx~600~$km~s$^{-1}$ in 
the direction of the constellation Hydra.

It is straightforward to show that the observation of Hubble's Law is consistent with what 
we would expect in a universe which is homogeneous (\ie~looks the same everywhere) and 
isotropic (\ie~looks the same in all directions). A number of modern cosmological observations 
support the properties of homogeneity and isotropy, including the distribution of galaxies 
on large scales (which will be the main topic of this thesis) and the smoothness of the 
cosmic background radiation. Together, assumptions of homogeneity and isotropy are known 
as the Cosmological Principle (CP).

\section{Our view of the Universe}
If one is to understand anything about the large scale structure of the universe, 
it is generally advisable to know where the galaxies that make up that structure 
are. Mapping and understanding the spatial galaxy distribution is a prerequisite 
for constructing a viable model of structure formation in the universe. This effort 
reached its first step with the Abell, Zwicky \& Lick catalogues, which eventually 
documented the angular positions of around a million galaxies. The second step then, 
was to expand the 2-D galaxy information by including the distances. This is distance 
to the galaxies via redshift surveys.

\subsection*{Abell, Zwicky \& Lick}
\begin{figure}
\centering
\includegraphics[scale=.8]{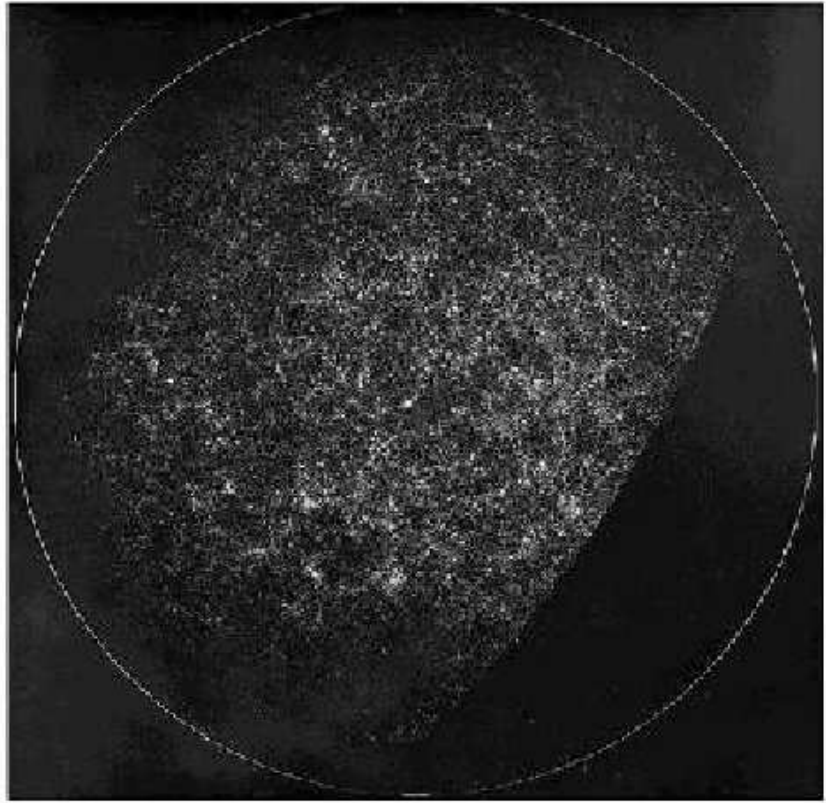}
\caption[Lick Galaxy Survey]{\small Lick galaxy survey, adapted from Peebles (1993)}
\label{fig:lick}
\end{figure}

Prior to the 80's, knowledge of the large scale structure of the universe was 
limited to only the angular distributions of galaxies, and a very uniform microwave 
background. In the late 50's, G. Abell (1958) collected several thousand angular 
positions of galaxies from the Palomar Sky Survey. This catalogue did not contain 
any information concerning the distance to the galaxies -- it was essentially a 
projection of the true galaxy distribution onto a sphere. Then in the 60's Fritz 
Zwicky and collaborators visually scanned thousands of photographic plates from the same survey, 
obtaining positions of over 30,000 galaxies in the northern sky (Zwicky\etal~1968). After that, in the early 
80's, Schechtman accumulated a catalogue of 1 million galaxies in the northern 
sky from the Lick astrographic survey, see figure \ref{fig:lick}.

\begin{figure}
\centering
\includegraphics[scale=1.]{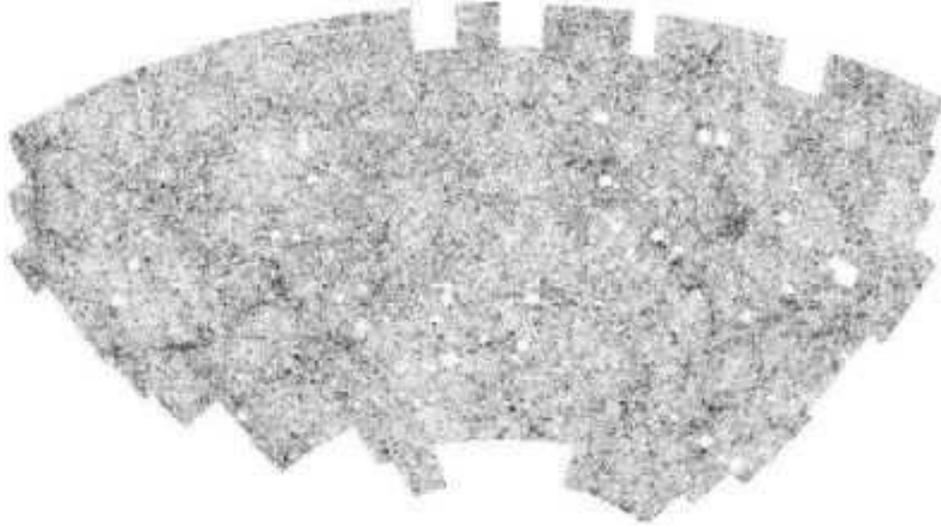}
\caption[The APM Galaxy Survey]{The APM galaxy survey. Maddox \etal~1990.}
\label{fig:apm}
\end{figure}

The Cambridge APM survey followed in the early 90's cataloguing about 2 million galaxies in the 
Southern Galactic Cap. Maddox \etal~1990. See figure \ref{fig:apm}

\subsection*{Redshift Surveys}
Redshift measurements involve determining the spectrum of the object to be measured. Once that 
is known, recognisable spectral lines can be found, and their deviation from their normal positions 
used to find the object's redshift. The Hubble Law then allows one to turn that redshift into a 
radial distance from our galaxy.

When redshifts were first being measured, it would typically take a few 
hours on a large telescope to collect enough photons to obtain the required spectrum. Once 
telescopes with enough light gathering power became available and spectroscopic detectors 
became sophisticated enough to allow many redshifts to be taken simultaneously in a reasonable amount of 
time, astronomers started using these instruments to make maps of the 3 dimensional locations 
of galaxies and galaxy clusters.

\begin{figure}
\centering
\includegraphics[scale=.75]{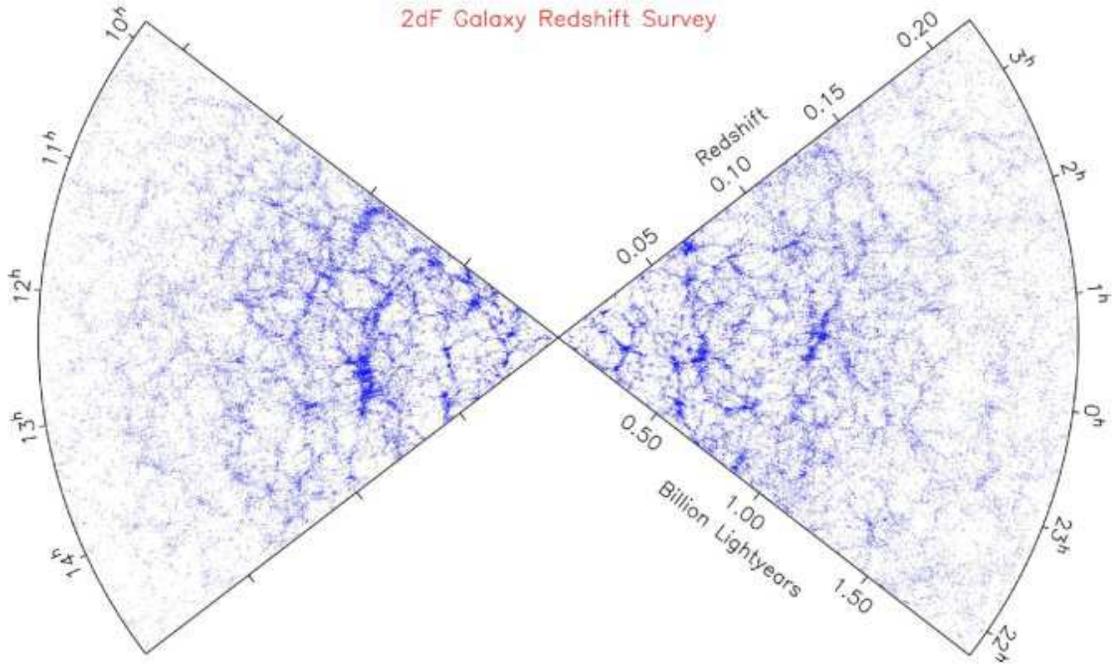}
\caption[The 2dF galaxy redshift cone]
	{\small 2dF galaxy redshift survey.}
\label{fig:2df_cone}
\end{figure}

Figure \ref{fig:2df_cone} is a  representation of some of the measured 3 dimensional galaxy 
positions in redshift space. The radial coordinate in this plot is the measured redshift 
(essentially indicating the distance from us) and the angular coordinates represent the 
angular position of the objects in the sky.  Thus the 3-D mapping of the universe began: 
\begin{itemize}
\item{} In the late 70's red-shift surveys finally became a reality with the very successful CfA 
(Center for Astrophysics) survey (Huchra \etal~1983). They managed to record 1100 spectroscopic 
red-shifts. 
\item{} In the early 90's the PSCz red-shift survey (Saunders \etal~2000) mapped 15,000 spiral 
galaxies from $\sim 83\%$ of the sky. This still stands as the largest survey in terms of sky coverage.
\item{} From 1998 to 2003, the Two Degree Field Galaxy Redshift Survey, using the 
Anglo-Australian telescope, accumulated 220,000 galaxy red-shifts. See Colless \etal~1999. 
This survey is illustrated in figure \ref{fig:2df_cone}.
\item{} The Sloan Digital Sky Survey (SDSS) is an ongoing attempt to collect 1 million 
galaxy red-shifts. The current release, DR5, contains 674,749 galaxies. See Percival \etal~2006 (DR5) and 
Abazajian \etal~2005 (DR4).
\end{itemize}
 
\subsection{Problems with this view}\label{problems}
The information directly available to us on the observed spatial distribution of galaxies is 
systematically biased compared to the true galaxy distribution.

Angular positions are trivial, but precise distance estimates are more difficult to obtain. 
More serious, however, is the problem that the sampling rate of observable galaxies depends 
strongly on redshift. If one considers the universe to be homogeneously distributed with 
matter, then we would expect the number of objects, in some observed solid angle of the sky, 
to grow as $r^{3}$. However this is not what we actually observe, instead we see 
distributions like figure (\ref{fig:histogram}), where the reference curve 
initially has the form of $r^{3}$ but soon drops off. This effect is due the 
diminished flux of faraway objects and our inability to collect enough photons 
from them to make valid observations. This `selection effect' can be visualised in figure 
(\ref{fig:2df_cone}), as the fall off in the apparent number density of sampled galaxies 
at larger redshift. This is accounted for by weighting each observed object with the 
inverse of the {\em selection function}, $\phi(r)$, this increases the contribution of 
counted objects. The functional form of the {\em selection function} is,
\be 
\phi(r) = 
\left( {r_{\circ}}\over{r} \right)^{2\alpha}
\left (
{r_{\star}^2 + r_{\circ}^2}
\over
{ r_{\star}^2+ {r}^2} 
\right) ^{\beta}.
\ee
In the above expression $r_{\star}, \alpha, \beta$ and $r_{0}$ are set parameters and 
$r$ is the proper distance to the observed galaxy. 

Apart from the red-shifts, distances can also be estimated from the flux, $F_{\upsilon}$, received from
a source. To accurately use this method, the intrinsic luminosity, $L_{\upsilon}$, of the source must be 
well known and that it radiates in a particular fashion, \ie~ beamed or uniform. Considering the 
source to radiate uniformly, the distance $d$ can be calculated from,
\be
d^{2}_{L}=\frac{L_{\upsilon}}{4{\pi}F_{\upsilon}}.
\label{luminosity}
\ee
The main problem with this distance estimator is that the vacuum of space is not so empty. It is 
filled with dust particles and gas, collectively termed the Interstellar Medium (ISM). The ISM can cause 
extinction of light (diminished flux) and leads to eq.(\ref{luminosity}) giving the wrong answer. The 
extinction of light is maximal at low galactic latitudes due to the dust content of our own galaxy. In fact 
at its worst this effect can completely block out the light coming from distant sources. 

In figure \ref{aitoff}, the masked regions (black) are shown for the \pscz~survey. The majority of it 
is due to the extinction of light through the galaxy. The sweeping arcs (north and south) are due to 
cryogenic problems in the satellite near the end of the survey period, it was not completed. In 
figure \ref{fig:masks}, some masked regions for the 2dF survey are shown. These masks are placed 
over the survey because local object may be blocking the view in a particular direction.

\chapter{The Universe at Large}
\vspace{-0.6cm}
Of the four fundamental forces of nature, the universe on large scales is governed by a 
single force: Gravity. In the proceeding section 
we will employ the use of Einstein's General Relativity to construct the main equations 
of cosmological evolution. From there we will relax our use of complicated tensor algebra 
and work in a Newtonian approximation to study the departures from the homogeneous Friedmann 
equations, due to the growth under gravity of tiny density inhomogeneities in the Universe. 
This is the `first order' Universe.

\section{General Relativity}
In 1915 Einstein developed the theory of General Relativity to explain gravity as a consequence 
of the fundamental connection between the geometry and matter content of space-time. This can be 
summed up in the neat phrase `Space-time tells matter how to move and matter tells space-time how 
to curve'. More specifically, Einstein developed a set of equations which balanced the curvature 
of space-time and the matter it contained. It is this curvature which generates the force of 
gravity and determines how matter moves within it. Generally, this can be written as,
\begin{equation}
8\pi GT_{i k} = G_{ik} = R_{ik}-\frac{1}{2}g_{ik}R-\Lambda g_{ik}.
\label{eq:einstein}
\end{equation}
To explain the ideas and physical motivation of the above equation is beyond the scope 
of this thesis (for further details see Misner \etal~1973 and Weinberg 1972 for a more 
physical description). In eq.(\ref{eq:einstein}), $G_{ik}$ is the Einstein tensor, 
$R_{ik}$ and $R$ are the Ricci tensor and scalar respectively. $g_{ik}$ is the space-time 
metric and $\Lambda$ is the Cosmological Constant. Although its inclusion in 
eq.(\ref{eq:einstein}) was described by Einstein to be his ``Greatest blunder'', it is 
nowadays a very crucial parameter in cosmology. We will describe in more detail each of 
the above parameters as we go along.

In our study of cosmology we need a theory of gravity, which we have, and a metric 
to describe the space-time. A metric is a function that measures the distance between 
events in space-time. Assuming that the CP is true, the form of this metric must conserve 
the CP as described in \S\ref{CP}; \ie~it must be homogeneous and isotropic. To construct 
the metric with the CP in mind, we slice up the 4-D space-time along $x^{0}=const.$ 
hypersurfaces, \ie~constant time.

To impose isotropy it is easier to tackle this problem in spherical coordinates since 
there is no preferred direction. For homogeneity we insist that the Ricci scalar is constant 
at all points on the hypersphere. And since we are dealing with a universe which is expanding 
it would be to our advantage to use a coordinate system which reflects this property. So 
we will work in a comoving system and therefore preserve the positions of galaxies relative 
to one another. Under the above assumptions it can then be shown (e.g. Carroll S.M, 2004) 
that the metric takes the following form, 
\begin{equation}
ds^{2}=-dt^{2}+a^{2}(t)\left [\frac{dr^{2}}{1-kr^{2}}+r^{2}(d\theta^{2}+\sin^{2}\theta d\phi^{2})\right ].
\label{eq:RWmetric}
\end{equation}
This is known as the Roberson-Walker metric (hereafter RWM). 
In the above expression $a(t)$ is the cosmic scale factor, $k$ is a constant which determines 
the spatial curvature and we are working in units where $c=1$. $k$ may take any real value but 
by suitable re-scaling of our coordinates it is only necessary to consider 3 possibilities.
\end{spacing}
\begin{spacing}{1.0}
\begin{itemize}
\item $k = 0$, corresponds to zero curvature and thus a flat universe. Initially parallel 
trajectories remain parallel.
\item $k = +1$, has positive curvature and the universe is closed. Its geometry is like the 
surface of a sphere. Initially parallel trajectories will converge.
\item $k = -1$, has negative curvature and leads to an open universe. Its geometry can be 
represented by the surface of a saddle. Initially parallel 
trajectories will diverge.
\end{itemize}
The RWM in component notation is,
\begin{equation}
g_{\alpha \beta} = 
\begin{pmatrix}
 -1 & 0 & 0 & 0\\
  0 & {a^{2}(t)}/(1-kr^{2}) & 0 & 0\\
  0 & 0 & a^{2}(t)r^{2} & 0\\
  0 & 0 & 0 & a^{2}(t)r^{2}\sin^{2}\theta \\
\end{pmatrix}.
\end{equation}
\end{spacing}
\begin{spacing}{1.5}
With the above information about the metric we can go about obtaining the non-vanishing 
components of the Christoffel Symbols, $\Gamma^{\gamma}_{\beta \mu}$. The components are 
calculated via,
\begin{equation}
\Gamma^{\gamma}_{\beta \mu} = \frac{1}{2}g^{\alpha \gamma}(g_{\alpha \beta,\mu} + 
g_{\alpha \mu,\beta} + g_{\beta \mu,\alpha}),
\label{eq:christoffel}
\end{equation}
where $\alpha$ is a summation index and a comma denotes partial derivative. Now it is 
merely a case of turning the handle to derive the essential non-zero elements of the 
Christoffel Symbols for the metric given by eq.(\ref{eq:RWmetric}),
\begin{equation}
\begin{aligned}
\Gamma^{0}_{11} & = \frac{a \dot{a}}{1-kr^{2}}\\
\Gamma^{0}_{22} & = a \dot{a} r^{2}\\
\Gamma^{1}_{22} & = r(kr^{2}-1)\\
\Gamma^{2}_{33} & = -\sin\theta \cos\theta\\
\Gamma^{1}_{01} & =\Gamma^{2}_{02}=\Gamma^{3}_{03}=\frac{\dot{a}}{a}
\end{aligned}
\qquad\text{}\qquad
\begin{aligned}
\Gamma^{1}_{11} & = \frac{kr}{1-kr^{2}}\\
\Gamma^{0}_{33} & = a \dot{a} r^{2}\sin^{2}\theta\\
\Gamma^{1}_{33} & = r(kr^{2}-1)\sin^{2}\theta\\
\Gamma^{3}_{23} & = \cot\theta\\
\Gamma^{2}_{12} & =\Gamma^{3}_{13}=\frac{1}{r}
\end{aligned}
\end{equation}
From the above set of equations we can now determine the nonzero components of the Ricci tensor, 
$R_{\alpha \beta}$.This is constructed by contraction of the Riemann tensor, 
$R_{\mu \nu \gamma}^{\lambda}$.
\begin{equation}
R_{\alpha \beta} \equiv R^{\lambda}_{\alpha \lambda \beta}.
\end{equation}
Summation is implied on the index $\lambda$ according to the {\em Einstein summation convention}. 
The components of the Ricci tensor are therefore related to the Christoffel symbols, above by,
\begin{equation}
R_{\alpha \beta}
=\Gamma_{\alpha \beta}^{\tau}\Gamma_{\tau \rho}^{\rho}
-\Gamma_{\alpha \rho}^{\tau}\Gamma_{\tau \beta}^{\rho}
+\Gamma_{\alpha \beta,\rho}^{\rho}
-\Gamma_{\alpha \rho,\beta}^{\rho}.
\end{equation}
In the above expression summation is implied over $\tau$ and $\rho$.
As an example this leads to, for the $tt$ component:
\begin{equation}
\begin{aligned}
R_{t t}
&  =\Gamma_{t t}^{\tau}\Gamma_{\tau \rho}^{\rho}
-\Gamma_{t \rho}^{\tau}\Gamma_{\tau t}^{\rho}
+\Gamma_{t t,\rho}^{\rho}
-\Gamma_{t \rho,t}^{\rho} \\
&  = 0 
-\Gamma_{t \rho}^{\tau}\Gamma_{\tau t}^{\rho}
+ 0 
-\Gamma_{t \rho,t}^{\rho} \\
&  = -3\left (\frac{\dot{a}}{a}\right )^{2}
-3 \frac{d}{dt}\left (\frac{\dot{a}}{a}\right ) \\
&  = -3\frac{\ddot{a}}{a}.
\end{aligned}
\end{equation}
The rest of the nonzero components are:
\begin{equation}
\begin{aligned}
R_{r r} & = \frac{a\dot{a}+2\dot{a}^{2}+2k}{1-kr^{2}}, \\
R_{\theta \theta} & = r^{2}(a\dot{a}+2\dot{a}^{2}+2k), \\
R_{\phi \phi} & = r^{2}(a\dot{a}+2\dot{a}^{2}+2k)\sin^{2}\theta.
\end{aligned}
\end{equation}
The Ricci scalar, $R$, is then obtained via the contraction of $R_{\alpha \beta}$ i.e,
\begin{equation}
\begin{aligned}
R & = g^{\alpha \beta}R_{\alpha \beta}.
\end{aligned}
\end{equation}
Therefore,
\begin{equation}
\begin{aligned}
R & = g^{t t}R_{t t} + g^{r r}R_{r r}
+ g^{\theta \theta}R_{\theta \theta}
+ g^{\phi \phi}R_{\phi \phi}, \\
& = 6\left [\left (\frac{\dot{a}}{a}\right )^{2}+\frac{\ddot{a}}{a}+\frac{k}{a^{2}}\right ].
\label{eq:ricci_scalar}
\end{aligned}
\end{equation}
Now looking back at equation (\ref{eq:einstein}), we can compute everything on the RHS of this equation. 
To evaluate the LHS we must define our energy-momentum tensor, $T_{i k}$. This tensor describes 
the matter and energy content of the universe. A perfect fluid approximation is completely defined 
by two quantities: the rest frame energy density $\rho$ and the isotropic rest frame pressure $p$.
\end{spacing}
\begin{spacing}{1.0}
\begin{equation}
T_{i k} = 
\begin{pmatrix}
 \rho & 0 & 0 & 0\\
  0 & p & 0 & 0\\
  0 & 0 & p & 0\\
  0 & 0 & 0 & p\\
\end{pmatrix}.
\label{eq:energy-momentum}
\end{equation}
\end{spacing}
\begin{spacing}{1.5}
Now using eq.(\ref{eq:energy-momentum}) with eq.(\ref{eq:einstein}) and assuming that $\Lambda = 0$, 
we can obtain two different equations. From the $tt$ component we get,
\begin{equation}
\left(\frac{\dot{a}}{a}\right)^{2} = H^{2} = \frac{8 \pi G}{3}\rho - \frac{k}{a^{2}},
\label{eq:fried1}
\end{equation}
and from the other components we obtain
\begin{equation}
\left(\frac{\ddot{a}}{a}\right) = -\frac{4 \pi G}{3}\left(\rho + 3p \right).
\label{eq:fried2}
\end{equation}
These expressions are known as the Friedmann equations. $H$ is the Hubble parameter and is defined 
as,
\begin{equation}
H = \frac{\dot{a}}{a}.
\label{eq:hubble}
\end{equation}
Apart from the global expansion, we can now investigate other aspects of the RWM universe. 
The {\em density parameter}, $\Omega$ is defined as,
\begin{equation}
\Omega = \frac{8 \pi G}{3H^{2}}\rho \equiv \frac{\rho}{\rho_{c}},
\end{equation}
with $\rho_{c}$ being the critical density required to produce a flat universe. Combining the 
above equation with (\ref{eq:fried1}) gives,
\begin{equation}
H^{2}=\frac{8\pi{G}}{3}\rho_{c}\Omega-\frac{k}{a^{2}}=H^{2}\Omega-\frac{k}{a^{2}},
\end{equation}
and rearranging this gives,
\begin{equation}
\Omega-1=\frac{k}{a^{2}H^{2}}.
\end{equation}
The special case of $\Omega = 1$ implies that $k=0$, and since $k$ is a fixed constant it must 
be concluded that $\Omega = 1$ for all time.

\section{Structure Formation}
There are two crucial observations that any model of structure
formation has to explain: the quadrupole anisotropy of CMB, as measured
by WMAP, is one part in 10$^5$ (Hinshaw \etal~2003) suggesting that the amplitude 
of the fluctuations was very small at the epoch of recombination; while redshift 
surveys of the Local universe show highly inhomogeneous matter distributions over 
galactic and cluster scales. The {\it  gravitational instability} is believed to be 
the physical mechanism which amplifies the small primeval fluctuations into the
structure that we observe today.

We now have the main ingredients for analysing the evolution of structure, at least within a 
linear approximation, these are contained within equations (\ref{eq:fried1})-(\ref{eq:hubble}). 
Combining the Friedmann equations and rearranging gives us the {\em continuity} equation,
\begin{equation}
\dot{\rho} = -3H\left(\rho + p \right),
\label{eq:continuity}
\end{equation}
which shows the conservation of energy as the universe expands. To proceed we will also make the 
assumption that the matter content of the universe is well described by an ideal gas equation of 
state, \ie ,
\begin{equation}
p = w\rho.
\label{eq:EOS}
\end{equation}
In standard cosmology we assume that the majority of the matter is in the form of cold dark 
matter (CDM), which has the equation of state, $w=0$. So the pressure component is assumed 
to be negligible which greatly simplifies the following calculations. 

In eq.(\ref{eq:RWmetric}) the effect of the curvature is small for distances much less 
than the Hubble radius $cH_0^{-1}= 3000 \hm$ (a variety of observations clearly favour 
$\Omega _{tot} \approx 1$ and $\Omega_{\Lambda} \approx 0.7$ therefore $|k|<H^2_0$). 
Hence, the RWM is well approximated by $ds^2= cdt^2-a^2(t)(dx^2+dy^2+dz^2)$~
(similar to the flat Minkowski metric of special relativity but with a time varying 
rescaling.), where $(x,y,z)$ denote the comoving Cartesian coordinates.  Assuming a 
conformal Newtonian gauge\footnote{Also known as the longitudinal gauge. This transform 
is very useful when considering scalar perturbations.} (\cf~Ma \& Bertschinger 1995) 
the Einstein field equations applied to the first-order perturbations $\phi$ of such a 
metric yield the Poisson equation of Newtonian gravity:
\begin{equation}
\nabla^2 \phi = 4\pi a^2 G \delta {\bar{\rho}},
\label{eq:poissongr}
\end{equation}
where $\delta {\bar{\rho}} \equiv \rho(\xvec,t)-{\bar \rho}(t)$ indicates the fluctuation 
of the mass density about the mean density $\bar{\rho}(t)$ and $\phi$ is interpreted as 
the Newtonian potential. Note that eq.(\ref{eq:poissongr}) does not assume that 
$\delta \bar\rho$ is small.

\subsection{The Eulerian Formalism}
\label{subsection:eulerian-formalism}
The Eulerian Formalism considers the large scale universe as a 
continually expanding fluid, whereby momentum, energy and mass 
conservations are encapsulated within the equations below:
\be
a{\dot{\delta}}+ \nabla \cdot [(1+\delta) {\vvec}]=0
\label{eq:continuityeuler}
\ee
\be
 a {\dot{\vvec}}+ \left  ( {\vvec} \cdot \nabla \right ) {\vvec} +
 {\dot a} \vvec= -\nabla \phi - \bar{\rho}^{-1} \nabla p,
\label{eq:eulereuler}
\ee
\be
\nabla^2 \phi = 4\pi a^2 G \delta \bar\rho,
\label{eq:poissoneuler}
\ee
Equations (\ref{eq:continuityeuler})-(\ref{eq:poissoneuler}) are the
equations of motion of a non-relativistic perfect fluid in comoving
coordinates. Eq.(\ref{eq:continuityeuler}) is the continuity equation
(expressing mass continuity) and eq.~(\ref{eq:eulereuler}) is the
Euler equation (conservation of the linear momentum).  In this system
of differential equations the over-density field, $\delta(\xvec,t)$,
appears rather than the usual density field,
$$\rho(\xvec,t) \equiv
\bar{\rho}(t)[1+\delta(\xvec,t)],
$$
with $\bar{\rho}$ being the spatially averaged mean density and the 
peculiar velocity, $\vvec(\xvec,t)$, is defined as 
\begin{equation}
\vvec \equiv
\frac{d\rvec}{dt}-\frac{{\dot {a}}}{a} \rvec.
\label{eq:pec_vel_def1}
\end{equation}
Over-dots indicate partial time derivatives. The pressure $p$ is related
to the density $\rho$ through equation (\ref{eq:EOS}). For
{\it adiabatic perturbations} there are no spatial variations in the
equation of state, therefore $\nabla p = w \nabla \rho = w
\bar{\rho}\nabla \delta$.

\subsubsection*{The Linear regime of Adiabatic perturbations}
Linearising our system of equation, we obtain
\begin{equation}
a\dot{\delta}+\nabla \cdot \vvec \approx 0,
\label{eq:lincontinuity}
\end{equation} 
\begin{equation}
a\dot{\vvec} + \dot {a}\vvec \approx  -\nabla \phi - c_s^2 \nabla \delta,
\label{eq:lineuler}
\end{equation}
\begin{equation}
\nabla ^2 \phi = 4\pi G a^2 \delta \bar\rho 
\label{eq:linpoisson}.
\end{equation}
Where $c_s$ represents the adiabatic sound speed, $c_s^2 \equiv
(\partial p / \partial \rho)_{s} = w$, and the subscript $S$ indicates
constant entropy throughout the space ($\nabla S=0$).

A general vector field may be decomposed into a (potential)
longitudinal and a (rotational) transversal part:
\begin{equation}
\vvec(\xvec,t)= \vvec _{\|} + \vvec_{\bot},\qquad \nabla \times \vvec _{\|}= \nabla \cdot \vvec_{\bot} = 0,
\end{equation}
From the curl of eq.~(\ref{eq:lineuler}) it follows that 
\begin{equation}
\frac{\partial }{\partial t} {\left ( a\nabla \times \vvec \right)}=\frac{\partial }{\partial t} 
{ \left ( a \nabla \times \vvec _{\bot} \right )} = 0.
\end{equation}
This implies that rotational modes are not coupled to density
perturbations and decay as $a^{-1}$.  Combining the time derivative of
the linearised continuity with the divergence of the linearised Euler
(\cf~eq.~\ref{eq:lineuler}) we yield the equation of motion for the
longitudinal density perturbations
\begin{equation}
\ddot{\delta}+2\frac{\dot{a}}{a}\dot{\delta} = \left ( \frac{c_s }{a}
\right )^2 \nabla ^2 \delta  + 4\pi G  \delta \rho.
\label{eq:jeansreal}
\end{equation}
Since the coefficients are spatially homogeneous (independent of
$\xvec$) this equation may be solved by expanding $\delta(\xvec,t)$ in
plane waves, $\delta(\xvec,t) = \delta_{\kvec}(t) e^{i\kvec\cdot
  \xvec}$, $\lambda = 2\pi a(t)/k$, where $\lambda$ is the proper
wavelength. After some straightforward calculations, it is easy to show
that the dynamical behaviour of $\delta_{\kvec}(t)$ obeys the following
differential equation:
\begin{equation}
{\ddot {\delta}} _{\kvec}+2 \frac{\dot{a}}{a} \dot{\delta}_{\kvec} = -c^2_s
\left ( k^2 - k_J^2 \right ) \delta _{\kvec},
\label{eq:jeanslinear}
\end{equation}
where we have defined the comoving Jeans wavenumber $k_J$ by
\begin{equation}
k_{J}= a\left( \frac{4\pi G\bar{\rho}}{c_s^2 } \right )^{1/2}.
\end{equation}
Two qualitative behaviours of the solutions can be easily discerned from
eq.~(\ref{eq:jeanslinear}). For wavenumbers larger than $k_{J}$
pressure dominates the right hand term and perturbations do not grow,
merely oscillate. For $k < k_J$ self-gravity dominates so that
gravitational instability can take place. Exact solutions to
eq.~(\ref{eq:jeanslinear}) exist for a variety of cases (see, for
instance, Peebles 1980). Since the dynamical
behaviour of ${\delta}_{\kvec}(t)$ is governed by a second order
differential equation, in general, there is one monotonically growing
solution and one monotonically decaying solution. In the limit $k \ll k
_J$ the effects of the pressure $p$ are negligible and thus all modes
grow at the same rate. In this regime, the general solution to
eq.~(\ref{eq:jeansreal}) is given
\begin{equation}
\delta(t,\xvec)= A(\xvec)D_{_{+}}(t) + B(\xvec)D_{_{-}}(t)
\approx   A(\xvec)D_{_{+}}(t)
\label{eq:general_solution1}
\end{equation}
where $D_{_{+}}(t)$ and $D_{_{-}}(t)$ are the growing and decaying
modes, respectively, while $A(\xvec)$ and $B(\xvec)$ are time
independent functions (Heath 1977).  The decaying
solution is a perturbation with initial over-density and peculiar
velocity arranged so its initial velocity quickly becomes negligible
(Peebles 1980). Thus for most of the history of
the universe the growing solution quickly comes to dominate. In an
Einstein-de Sitter universe $D_{_{+}}(t)\propto a$ and
$D_{_{-}}(t)\propto a^{-3/2}$. For a dust universe with $\Omega$ $<$1,
the growing mode $D_{_{+}}$~is
\begin{equation}
D_{_{+}}(t)=\frac{3\sinh \eta (\sinh \eta - \eta)}{(\cosh \eta - \eta)^2}-2.
\end{equation}
$\eta$ indicates conformal time $\eta \equiv (-k)^{1/2}\int^tdt'/a(t')$.

Given a solution for the density perturbation field $\delta(\xvec,t)$,
the velocity, gravitational potential and gravity field follow. For the
longitudinal modes $\vvec = \vvec_{\|}=-\nabla \phi_v /a$. The gravity
field is $\gvec \equiv -\nabla \phi /a$. Thus, from the system of
differential equations, eq's.~(\ref{eq:lincontinuity})--(\ref{eq:linpoisson}), for $k\ll k_J$, we
obtain
\begin{equation}
\phi_v = \frac{a^2Hf}{4\pi}\int \frac{ \delta(\xvec')}{| \xvec
  ' - \xvec|} d^3 x',\qquad \phi = \frac{3}{2} \frac{\Omega
  H}{f}\phi_{\vvec},\qquad \gvec = \frac{3}{2}\frac{\Omega H}{f} \vvec. 
\end{equation}
Where $f$ is defined by
\begin{equation}
\label{eq:fdef}
f(\Omega,z)\equiv \frac{d\log D_{_{+}}}{d\log a}= \frac{\dot{D}_{_{+}}(t)}{H(t)D_{_{+}}(t)}.
\end{equation}
The behaviour of $f(\Omega,z)$ at the present epoch ($z$=0) is very
well described by $f \approx \Omega ^{0.6}$ in the case of universes
with negligible space curvature or rather small cosmological
constant~(Peebles 1980).

\section{$N$-body Simulations}
So far we have only been concerned with the linear regime. To tackle the 
non-linear regime we must leave our analytical tools behind and instead 
buy a very big computer to run $N$-body simulations. This is exactly what 
people have been doing since Sverre Aarseth wrote the first astrophysical 
$N$-body codes (Aarseth, 1978)\footnote{Aarseth helped greatly in the early 
days of $N$-body simulations by making his codes readily available and easily 
adaptable.}.

In the early days of this field, cosmological simulations could only handle roughly 1,000 
particles (Goth \& Turner, 1979). But with the exponential increase in computing power it 
was not long until very large $N$-body simulation were being carried out. By the mid-80's 
there were 30,000 particles being used (Efstathiou, 1985), this rose steeply to $10^{6}$ 
in the 90's (Bertschinger \& Gelb, 1991) and now the most recent effort by the ``Virgo 
Consortium'' has seen the first billion particle cosmological simulation (Evrard \etal~2002). 

Over the last two decades cosmological $N$-body simulations have played a crucial role in 
the study of the formation and evolution of cosmic structure. Primarily they have been used to 
match theory with observations. However, simulations like the recent ``millennium run'' 
(Springel \etal~2005) use only dark matter particles which form the gravitational potentials 
for structure growth. Therefore, there is a crucial step to go from dark matter particles 
to dark matter halos and finally individual galaxies. The basic technique for doing so is 
discussed in \S\ref{subsection:makingmocks}. One would hope that in this process we are not 
masking some underlying physics. 

The impressive progress achieved in the observational front with the completion of very large 
surveys such as 2dF and SDSS poses a clear challenge to the numerical work in cosmology: the 
precision of the predictions provided by the current $N$-body experiments have to be of the 
order of a few percent.

$N$-body simulations, however, do not include all relevant physics like magnetic fields and 
gas dynamics. They also have resolution issues, which must play a roll on small scales. 
So while modern simulations are undoubtedly very useful tools, they must be used with care.

\chapter{Quantifying The  Large Scale Structure}
\vspace{-0.6cm}
The issue of quantifying structure is not confined to cosmology, it is a complex matter 
which stretches across many areas of science. Patterns are there to be identified and 
exploited in order to find subtle connections and correlations between observables. This 
method of analysis can be seen in fields as diverse as studying the stock market and modelling  
biological systems. A wide variety of different statistical tools have been employed to quantify 
structure, but in studies of the large-scale distribution of galaxies perhaps the most common 
has been the 2-point correlation function

\section{The Two Point Correlation Function: $\xi$}\label{2PCF}
The two point correlation function (hereafter 2PCF), $\xi(r)$, is defined as the excess 
probability, with respect to a Poisson distribution, of finding two galaxies in volumes $dV_1$
and $dV_2$ separated by a distance $r$. The joint probability is then,
\begin{equation}
dP_{1,2}=n^{2}[1+\xi(r)]dV_{1}dV_{2},
\label{A}
\end{equation}
where $n$ is the average density of the sample. This is however not a straight forward 
working definition since we know that density is varying through the space. Consequently 
it is not straightforward to identify how large a volume of space must be sampled in order 
to reliably measure the average density of the galaxy population.  Equally importantly, 
even if such a `fair sample' volume can be identified, the number count of observable galaxies 
within that volume will not generally be a reliable estimate of the `true' number of galaxies 
in the volume because of observational selection effects - \ie~a radial selection function 
and angular masked regions.  Also, measuring accurately the distance of galaxies in a survey 
is not straightforward: redshifts are distorted by peculiar motions and redshift-independent 
distance indicators (\eg~using some form of `standard candle' assumption) are rather noisy.

For this work we proceed to measure the mean density $\bar{n}$ as the number of galaxies inside the 
survey, each weighted by the selection function, $\phi(r_{i})$; in practice we perform a 
{\em Monte Carlo} over the sample space (see Strauss \&~Willick, 1995 and references therein). 
We can also write the conditional probability, $dP_{c}$, of finding a galaxy in a volume $dV_1$, 
at a distance $r$ from another galaxy,
\begin{equation}
 dP_{c}=\bar{n}[1+\xi(r)]dV_1.
\label{B}
\end{equation}
From this equation it is straightforward to see the properties of the function $\xi(r)$. 
For $\xi(r)=0$, we recover a uniformly random point set, such that the probability of 
finding a galaxy in volume $dV$ is simply proportional to $dV$. The case of excess clustering is 
$\xi(r)>0$, where we have more galaxies than in the Poisson case. Then there is also 
the case where $\xi(r)<0$. This case corresponds to anti-clustering, and could be relevant 
\eg~to some models of galaxy formation where the formation of a galaxy may inhibit the 
formation of other galaxies in its vicinity.

In general the observed $\xi(r)$ is well described by a power law scaling 
with distance and it is standard practice in the literature to write the 2PCF as,
\begin{equation}
\xi(r)=\left (\frac{r_{0}}{r}\right )^{\gamma}.
\label{C}
\end{equation}
Here $r_{0}$ is a characteristic scale length usually evaluated when $\xi = 1$.
This description is however an over simplification as the physics of structure 
formation is richer in complexity than a simple power law.

One great advantage of using this estimator, is that the Fourier Transform of 
$\xi(r)$ gives the Power Spectra, ${\mathcal P}(k)$, of density anisotropies (see \S\ref{section:PS}). 
It is also a statistic which is very easy to visualise and very easy to compute. 
From $\xi(r)$ it is also possible to determine the correlation dimension, $D_{2}$, 
of the discrete point distribution. 
$D_{2}$ is calculated via,
\begin{equation}
D_{2}= 3+\frac{d [\log \xi(r)]}{d[\log (r)]}.
\label{eq:D2}
\end{equation}
For a homogeneous distribution in 3-D we would expect $D_{2} \rightarrow 3$. 
However, at scales around $r_{0}$, $\gamma \sim 1.77$ (Davis-Peebles, 1983), 
corresponding to  a dimensional value, $D_{2} \sim 1.23$. We will soon see where 
expression (\ref{eq:D2}) comes from when we discuss generalised dimensions in 
\S\ref{fractals}.

\subsection{$\xi$-Correlation Estimators}
There are a few different ways to measure $\xi(r)$, in the literature but the 
basic computational structure is more or less the same. The main differences 
between estimators are usually the way in which they deal with `edge effects' 
and the `shot noise'. We are obviously looking for correlations between galaxies 
so we begin by centering on a galaxy, and proceed to count the number of galaxies 
within spherical shells of different radii around the central object. This 
procedure is repeated by centering on all, or a randomly chosen subset of all, 
the galaxies in the catalogue to obtain a statistical average. The number of 
galaxies counted in each cell is then normalised by a Poisson term which is 
related to the mean density. This estimator is written as,
\be
\xi_{j}(r) = \frac{dP_{c}}{\bar{n}dV}-1 = \frac{DD}{DR}-1 
\ee\be
 = \displaystyle\sum_{i}^{N}\frac{\Theta(|r_{j}-r_{i}|-r).
\Theta(r+dr-|r_{j}-r_{i}|)}{\bar{n}(r).dV(r).\phi(r_{i}).\phi(r_{j})}-1,
\label{eq:my_2PCF}
\ee
where $DD$ are data-data pairs, $DR$ are data-random pairs and $N$ is the number of galaxies. 
$\phi$ is the selection function and 
\be
\Theta(x) = \left\{\begin{array}{cc} 0,& x < 0,\\
				       1,& x > 0,
\end{array}.\right.
\ee
The problem with eq.(\ref{eq:my_2PCF}) lies within the mean density term. To obtain $\bar{n}$, 
the most accurate method should be to sum over all the galaxies while weighting each by the 
inverse of the selection function. This is calculated as follows,
\begin{equation}
\bar{n} = \frac{1}{V}\sum_{i}^{N}\phi(r_{i})^{-1}.
\end{equation}
\begin{figure}
\centering
\includegraphics[scale=.7]{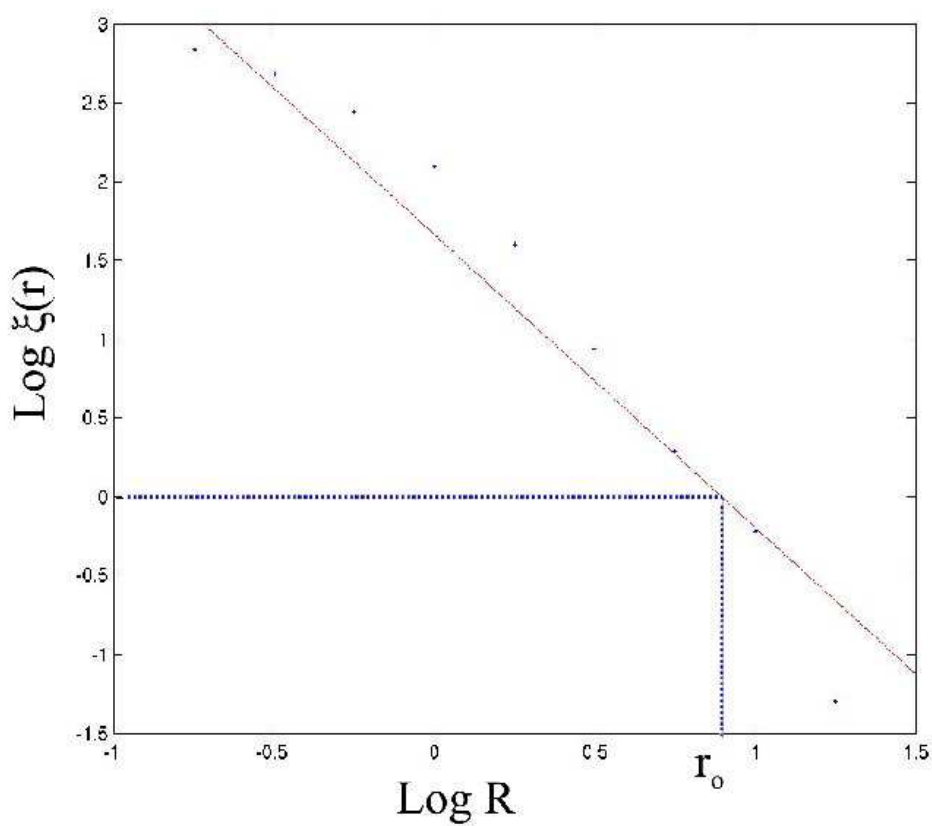}
\caption[The $\xi$-Correlation Function]{The 2PCF measured using (\ref{eq:my_2PCF}) and 
applied to the PSCz mock galaxy catalogue. The gradient gives $\gamma \approx 1.6$ and therefore 
$D_{2} \approx 1.4$ on scales up to 30Mpc}
\label{my_2PCF}
\end{figure}
In the last equation, $V$ is the volume and the sum is over all galaxies. To reduce the variance 
associated with $\xi(r)$, the average can be taken,
\begin{equation}
\langle\xi(r)\rangle=\frac{1}{N}\sum_{j}^{N}\xi_{j}(r).
\end{equation}
The angled brackets represent a statistical average, which usually invokes the use of the {\em 
cosmological ergodic theorem} (see \S\ref{ergodicity}). The result of applying eq.(\ref{eq:my_2PCF}) 
to a mock galaxy catalogue can be seen in figure \ref{my_2PCF}. The red line is a straight line fit 
to the data, giving $\gamma \approx 1.6$. 

So far we have used $DD$ and $DR$, however $RR$ pairs are also useful in calculating the 2PCF. 
Other estimators like the minimum variance estimator, by Landy \& Szalay (1993), use the 
$RR$ pairs to help correct for boundary effects. Their estimator takes the form,
\begin{equation}
\xi(r) = \frac{M(M-1)}{N(N-1)}\frac{DD}{RR}-\frac{(M-1)}{N}\frac{DR}{RR}+1,
\label{eq:szalay}
\end{equation}
where the $M$ is the number of random points. There are many more estimators for the 2PCF. 
Kerscher, Szapudi, \& Szalay (2000) show that equation 
(\ref{eq:szalay}) is strongly preferred over other methods. 

\subsection{Ergodicity}\label{ergodicity}
A short note on the Ergodic Principle:\\ 
The observed universe is unique.  This implies that averages have to
be spatial ones. Such averages will be equal to those obtained if 
instead we were to average over an ensemble of universes if the 
{\it Cosmic Ergodic Theorem} holds. {\it Ergodicity} in the cosmological 
context means that ensemble averaging and spatial averaging are equivalent. 
Note that, in contrast with the common practice in statistical mechanics, 
the cosmological Ergodic Hypothesis refers to the spatial distribution of
a random field at a fixed time rather than to the time evolution of
the system. Thus, for instance, the ensemble average of the random field
$\delta(\xvec)$ at a point $\xvec$, $\left < \delta(\xvec) \right >$,
is simply the expectation value of the random variable $\delta(\xvec)$.

\subsection{Gaussianity}
Let us introduce some of the statistics used by cosmologists to
characterise the spatial distribution of matter.  We define {\it
r.m.s.} $\sigma_{\delta}$ fluctuations of a continuous density field
$\delta(\xvec)$ as
\begin{equation}
\sigma^2_{\delta}\equiv \Big\langle \delta(\xvec)^2 \Big \rangle,
\end{equation} 
and the {\it correlation function} by
\begin{equation}
 \xi(r_{12}) \equiv \Big \langle \delta({\bf x_1}) \delta({\bf x_2}) \Big \rangle.
\end{equation}
(Note that for a homogeneous and isotropic random process $\xi$ only 
depends on the distance between the two points $r_{12}= |\xvec_1
-\xvec_2|$.) The correlation function is a measure of the spatial
correlation of the field $\delta (\xvec)$. 

A random field is said to be {\it Gaussian} if all $N$-point
multivariate probability distribution functions are multivariate
Gaussian distributions defined by their mean vector $\langle
\delta(\xvec_i)\rangle$ (which the ergodicity implies to be identically
zero) and their covariance matrix $M_{ij} = \xi(\xvec_i,\xvec_j)$.
Gaussianity is a very popular assumption for two reasons. The first one
is that the calculations are ``easy'' to perform. The second reason is that 
the CMB seems to support a Gaussian initial density field, Komatsu \etal~2003.

\subsection{Power spectrum}\label{section:PS}
If we expand the $\delta(\xvec)$ field in plane waves as 
\begin{equation}
\delta({\bf x}) = \frac{V_u}{(2 \pi)^{3}} \int e^{i{\bf k}\cdot {\bf
    x}}\delta_{\bf k}d^3 k,
\end{equation}
we see that its Fourier transform $\delta_{\kvec}$ is given by
\begin{equation}
  \delta_{\bf k}= \frac{1}{V_u}\int e^{-i{\bf k} \cdot {\bf x}}\delta({\bf x})d^3 x,
\end{equation}
where $V_u$ may be thought of a ``fair sample'' of the universe. 
The {\it power spectrum} of the density field $\delta(\xvec)$ is defined as the
expectation of the two-point function in Fourier space, as follows:
\begin{equation}
\Big \langle \delta _{\kvec_1}\delta _{\kvec_2} \Big \rangle \equiv P(k_1)
\delta_D(\kvec_1-\kvec_2)
\label{eq:powerspectra1}
\end{equation}
where $\delta_D$ is the well-known Dirac delta function. This implies
that even if $\delta_{\kvec}$ is not a Gaussian distribution, the
random variable $\delta(\xvec)$, being an infinite sum of independent
random variables, will still be Gaussian by the Central Limit Theorem
for some well-behaved power spectra.  We can see that the Dirac
function in eq.~(\ref{eq:powerspectra1}) is required because of the
translational invariance, $\langle \delta(\xvec_1)
\delta(\xvec_2)\rangle = \xi (| \xvec_1-\xvec_2|)$. Similarly, we can
also see that isotropy implies that $P(k)$ depends only the magnitude
of the wave-vector $\kvec$.

\subsection{Window Functions}
For some calculations it may be necessary to apply a cutoff at high
spatial frequencies, this is due to non-linearities on small scales. 
The smoothed field $\tilde{\delta}(\xvec)$ that may be obtained by convolution 
of the ``raw'' field with some weighting function $W$ (called {\it window function}) 
having a characteristic scale $r_W$ is given by
\begin{equation}
{\tilde{\delta}}(\xvec) \equiv \int \delta(\xvec')W(\xvec'-\xvec,r_{W})d^3x',
\label{eq:smoothed_dens}
\end{equation}
has {\it r.m.s.} $\sigma _{\tilde{\delta}}$~fluctuations given by
\begin{equation}
\sigma ^2_{\tilde{\delta}}= \langle | {\tilde{\delta}}(\xvec) |^2 \rangle =
  \frac{1}{(2\pi)^3 }\int |W(\kvec)| ^2 P(k) d^3k.
\label{eq:smoothed_densvariance}
\end{equation}
Where $W(\kvec)$ is the representation in Fourier space of
$W(\yvec,r_{W})$.  The window function has the following properties:
$W(\xvec'-\xvec,r_W) = \mbox{{\it const.}}\simeq r_W^{-3}$~if $|\xvec
-\xvec'|\ll r_W$, $W(\xvec'-\xvec,r_W) = 0$~if $|\xvec -\xvec'|\gg
r_W$, satisfying the relation $\int W(\xvec'-\xvec,r_W)d\yvec= 1$. One
of the most common window functions is the ``{\it top hat}'' (TH)
window function which is defined by the relation
\begin{equation}
W_{TH} (| \xvec - \xvec '|; r_{TH}) = \frac{3}{4\pi r_{TH}^3} H (
1 - \frac{|\xvec - \xvec'|}{r_{TH}}),
\end{equation}
where $H$ denotes the Heaviside step function $(H(y) = 0$ if $\, y \le 0,$
and $H(y) = 1 \,$ if $ y>0$). Another commonly used window function is
the {\it Gaussian} kernel:
\begin{equation}
W_{G} (| \xvec - \xvec'|; r_{G}) = \frac{1}{(2\pi r_{G}^2)^{2/3}}\exp 
\left ( - \frac{ |\xvec - \xvec'|^3}{2 r_{G}^2} \right ).   
\end{equation}

\section{Higher Order Correlations}
It is a natural question then to ask whether there are higher order correlations than 
the simple 2PCF, generally defined as, $\xi_{2}(r)$. The answer is most definitely 
YES. A Gaussian random field would in principle be completely defined by the 2PCF 
(the initial density field is thought to have this property), however due to non-linear 
structure formation we now have local non-gaussianity which means we must look to 
higher order moments to accurately quantify the galaxy distribution.

\begin{figure}
\begin{center}
\includegraphics[scale=.7]{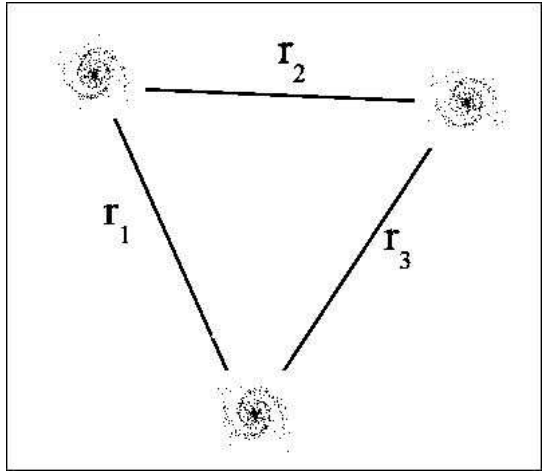}
\caption[Illustration of the 3 point correlation function]{\small The distances ${\bf r}_{1}$  
${\bf r}_{2}$ and ${\bf r}_{3}$ separated the three galaxies. The triangular configuration can 
be fixed at the outset or it could be included as a variable at the cost of much more computation.}
\label{fig:3PCF} 
\end{center}
\end{figure}

These higher order correlations are defined as, $\xi_{n}(r_{1},....,r_{n})$, where n is the 
order of the correlation function. As an example the 3 point correlation function is defined 
as the joint probability of there being a galaxy in volume elements $dV_{1}$ and $dV_{2}$ 
given that these elements have displacements ${\bf r}_{1}$  and ${\bf r}_{2}$ from the galaxy 
which is being investigated. This is illustrated in fig.(\ref{fig:3PCF}). The joint probability 
can be written as,
\begin{equation}
dP=n^{3}[1+\xi(r_{a})+\xi(r_{b})+\xi(r_{c})+\zeta(r_{a},r_{b},r_{c})]dV_{1}dV_{2}dV_{3},
\label{eq:3PCF}
\end{equation}
where $r_{a}$,$r_{b}$ and $r_{c}$ are the sides of the triangle. Assuming homogeneity and 
isotropy means leads to $\zeta$ being a symmetric function of the three lengths. Equation 
(\ref{eq:3PCF}) is the full three-point correlation function, and $\zeta$ is known as the 
reduced part. 

The conditional probability of finding two objects to complete the triangular configuration 
given that we are centring on a galaxy is,
\begin{equation}
dP=n^{2}[1+\xi(r_{a})+\xi(r_{b})+\xi(r_{c})+\zeta(r_{a},r_{b},r_{c})]dV_{2}dV_{3}.
\label{eq:3PCF_1}
\end{equation}
Then the conditional probability of finding a galaxy to complete the triangle, given that 
we have a pair of galaxies with separation $r_{a}$, is,
\begin{equation}
dP=n\frac{[1+\xi(r_{a})+\xi(r_{b})+\xi(r_{c})+\zeta(r_{a},r_{b},r_{c})]}{1+\xi(r_{a})}dV_{3},
\label{eq:3PCF_2}
\end{equation}

In extending to N-point correlation functions the computational load is increased exponentially. 
This limits our ability to determine accurately $N > 4$, however there may be a way to avoid this 
problem by making some approximations.

Given that $\xi(r)$ can be represented by a power law,
\be
\xi(r) = Br^{-\gamma}, \qquad \gamma \simeq 1.77,
\ee
the 3PCF is then found to be well described by a combination of $\xi$'s (Peebles, 1980), \ie,
\be
\zeta(r_{a},r_{b},r_{c}) = Q[\xi(r_{a})\xi(r_{b})+\xi(r_{b})\xi(r_{c})+\xi(r_{a})\xi(r_{c})],
\ee
with $Q \simeq 1.0 \pm 0.2$ (Meiksin, Szapudi \& Szalay, 1992).

\section{Minimal Spanning Trees}
One draw back with the 2PCF, as we have defined it in eq.(\ref{A}), is that it is insensitive to 
filamentary structure. This is due to it being a function only of distance and not direction; 
thus all angular information is lost through the averaging. The Universe does appear to contain 
filaments, walls and other such features, but whether these are real or due to chance alignments 
has of course been a topic of debate over recent decades.

To quantify this kind of structure Barrow, Bhavsar and Sonoda (1985) introduced a method 
from Graph Theory, called Minimal Spanning Trees (MST). The procedure to implement this 
is as follows:
\end{spacing}
\begin{spacing}{1.0}
\newcounter{Lcount}
\begin{list}{\small\arabic{Lcount}}
{\usecounter{Lcount}
\setlength{\rightmargin}{\leftmargin}}
\item A galaxy A is chosen as a starting point within a 3-D galaxy distribution. 
\item The nearest galaxy to A is labelled B and a straight line is drawn to connect 
the two. This line is known as a {\bf path}. 
\item The closest galaxy to the set of previous galaxies (in this case A and B) is 
added to the set and is connected to the closest galaxy in the set. This produces a 
branching behaviour, and this step is repeated. 
\item After some time we are left with many paths and {\bf circuits} (closed paths). 
If there are no circuits the graph is open and this is known as a {\bf tree}. 
\item To then transform this abstract visualisation into a numerical representation 
of structure we can do a few things, e.g.:
\newcounter{LLcount}
\begin{list}{\small\roman{LLcount}}
{\usecounter{LLcount}
\setlength{\rightmargin}{\leftmargin}}
\item Calculate the number of lines in angular bins of $d\theta$ w.r.t an adjoining line.
\item Calculate the number of lines in distance bins, dr. Obtaining dN vs dr. 
\end{list}
\end{list}

\begin{figure}
\begin{center}
\includegraphics[scale=0.65]{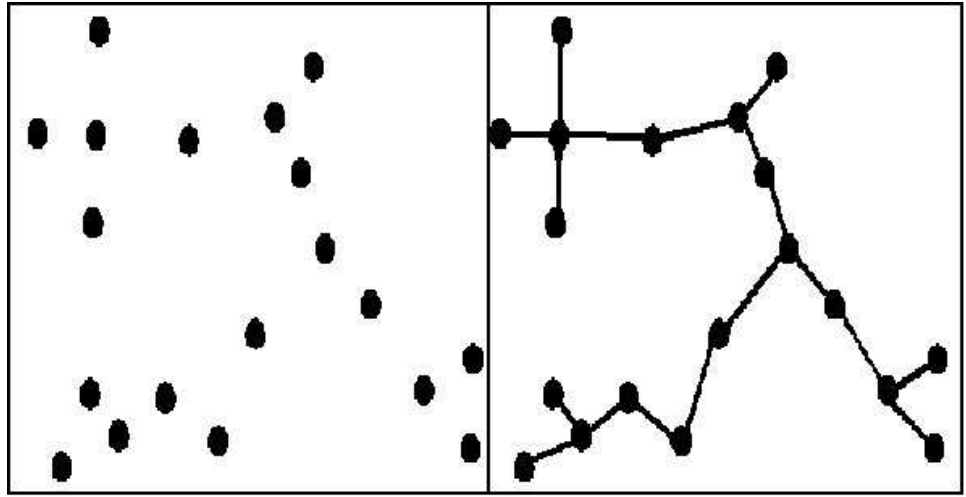}
\caption[Minimal Spanning Tree]
{\small On the left hand panel we see a distribution of points. As we apply the 
method from 1-5 the Minimal Spanning Tree is constructed on the right hand panel.}
\end{center}
\end{figure}
\end{spacing}
\begin{spacing}{1.5}
This method connects all galaxies in some fashion. But we know that not all galaxies are physically 
connected, as a galaxy in one cluster has little to do with another galaxy in a distant cluster.

To account for this, we adjust the previous method by only joining a galaxy to 
a pre-existing tree if its distance to the closest member is less than some threshold 
distance. This technique is known as separating and was introduced by Clark \& Miller (1966). 
This method was recently applied to the SDSS DR1 by Doroshkevich \etal (2004). They found that 
groups and clusters are more likely to be found close to walls rather than filamentary structure.

\section{The Fractal Universe}\label{fractals}   

Fractal patterns can be thought of as the place where chaos and order meet. This is because 
self-similarity (fractality) seems to be an eventual by-product of chaotic systems. 
Fractal pictures are usually associated with ``Julia'' and ``Mandelbrot'' sets named after the 
French mathematicians. 
\begin{figure}[h]
\centering
\includegraphics[width=6cm, height=4cm]{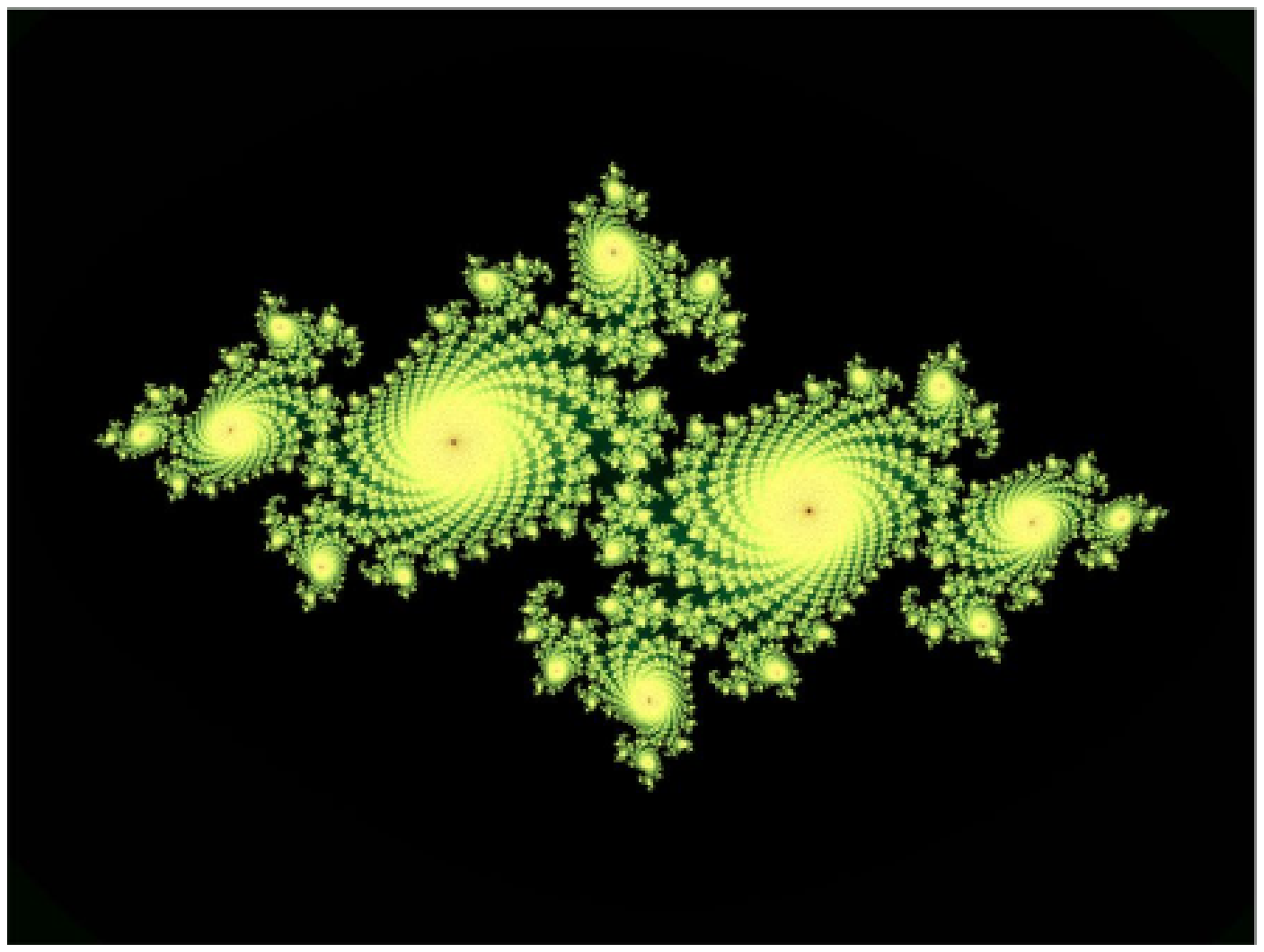}
\includegraphics[width=6cm, height=4cm]{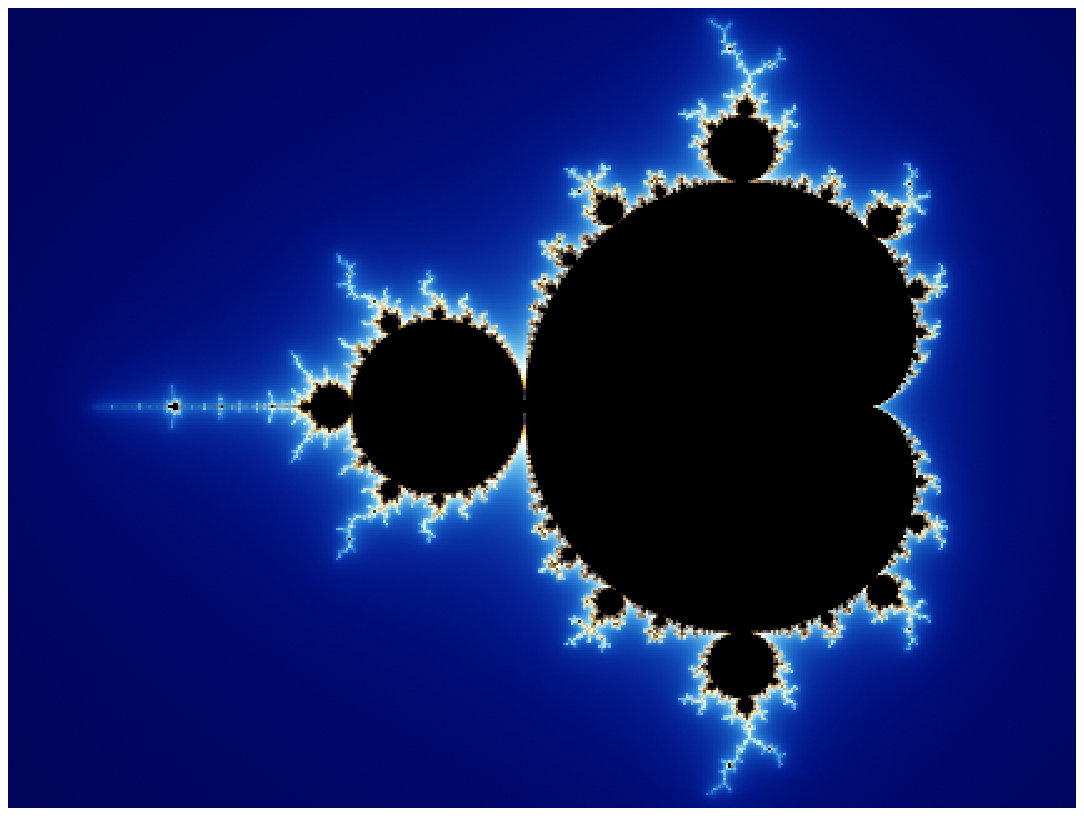}
\label{fig:fractals}
\caption[Fractals: Julia \& Mandelbrot sets]
{\small {\em Left:} The Julia set. {\em Right:} The Mandelbrot set. 
In both the Julia \& the Mandelbrot sets the self-similarity is clearly apparent \ie~successive 
enlargements of areas will show similar patterns to the picture as a whole. }
\end{figure}

Fractals belong to a branch of mathematics known as Fractal Geometry. Unlike the usual 
definition of Geometry, where regular shapes and patterns are studied, fractals can be 
highly irregular and the dimensions which are explored are not confined to integer values like 
$D=2$ for surfaces and $D=3$ for volumes. The term fractal was originally coined by Mandelbrot himself, who is 
widely considered the father of fractal studies. In fig.(\ref{fig:fractals}) the Mandelbrot
and Julia sets are shown. These are classical examples of fractal patterns, we can easily 
see that zooming in on certain areas will yield patterns which are reproductions of the 
whole picture. It is this self-similarity in fractal patterns which make them 
scale-invariant objects. Multifractals on the other hand are not scale invariant, since 
their spatial dimension can vary with scale length. In these terms monofractals can be 
thought of as a special case of a more general multifractal.

Fractals were merged with the physical sciences through their intimate connection to 
nonlinear physics and chaotic dynamics. In configuration space we may have chaotic motion for 
an unstable system, however, in phase space dynamics may be more ordered and the system may evolve 
towards an attractor.  This is rather like a chaotic analogy to an equilibrium state.  Phase space 
trajectories are, however, by no means an easy-to-use tool for analysing structures since, in 
cosmology, we only have access to spatial coordinates\footnote{This is not entirely true as we 
do have limited velocity information as well.}. Nevertheless, this kind of ordered phase space 
gives rise to a self similar pattern in real space. This is, more specifically, a fractal pattern. 

Now it is known that on large scales, the main ingredient to structure formation 
is the force of gravity. However the $1/r$ potential for gravity leads to highly 
nonlinear motions and also to cross talking between different spatial scales. So it 
is not so great a jump to consider the galaxy distribution as a fractal of some 
kind. In fact this line of analysis is not a new one. The distribution of galaxies 
in the universe has already been shown to be well described using a multifractal 
framework, see \eg~Jones \etal~1988. 

\subsection{Multifractal Formalism}\label{MFA}
In this analysis we will adopt the procedure layed out in Henschel \etal~(1983) 
to determine the R\'{e}nyi (Generalised) dimensions of a
point set embedded in a three-dimensional Euclidean space. The
probability of a galaxy, $j$, being within a sphere of radius $r$
centred on galaxy $i$ is,
\begin{equation}
\begin{aligned}
{p}_{i}(r) & =\frac{{n}_{i}(r)}{{N}}, \\ 
           & =\frac{1}{{N}}\sum_{j=1}^{N}\Phi(|r_{i}-r_{j}|-r).
\label{prob}
\end{aligned}
\end{equation}
Here $n_{i}(r)$ is the number of galaxies within radius $r$, $N$ is the total number 
of galaxies and 
\begin{equation}
\Phi(x) = \left\{\begin{array}{cc} 1,& x < 0,\\
				       0,& x > 0,
\end{array}\right.
\end{equation}
Equation (\ref{prob}) can then be related to the partition sum via Grassberger and 
Procaccia's (1983) correlation algorithm,
\begin{equation}
Z(q,r)=\frac{1}{M}\sum_{i=1}^{M}[{p}_{i}(r)]^{q-1} \propto r^{\tau (q)}.
\label{partition}
\end{equation}
In this case $M$ is the number of counting spheres and $q$ defines the
generalised dimension we are investigating. $\tau (q)$ is the
scaling exponent, which is then related to the infinite set of
dimensions through,
\begin{equation}
D_{q}=\frac{\tau (q)}{q-1}, \ \ \ \ q\neq 1.
\label{eq:dimension}
\end{equation}
Clearly the special case of $q=1$, the information dimension, cannot
be determined using the above expression but can be found
approximately in the limit $q\rightarrow 1$. This is an important
dimension to calculate as it gives equal weighting to voids and
clusters. Voids are enhanced for $q < 1$ and clusters are enhanced
for $q > 1$, so $q=1$ is in this sense the most unbiased dimension 
in the set. To determine $D_{1}$ more accurately we must calculate,
\begin{equation}
S(r)=\frac{1}{M} \sum^{M}_{i=1}{\rm log}\,[p_i(r)] \propto r^{D_{1}},
\end{equation}
where $S(r)$ is the {\it partition entropy} of the point set.

\begin{figure}
\centering
\includegraphics[width=6in]{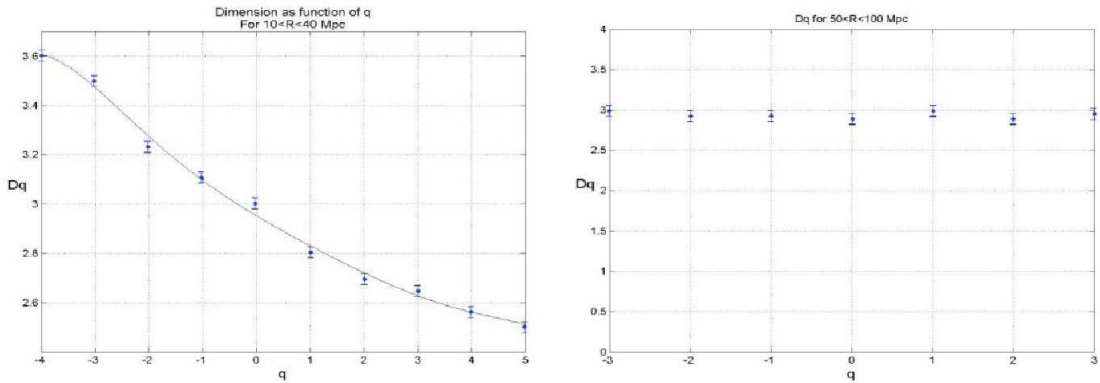}
\caption[$D_{q}$ curves for simulated halo catalogue.]
{\small Applying the multifractal analysis to $\Lambda$CDM halo catalogue, we obtain the usual $D_{q}$ 
curve. A linear $\chi^{2}$ fit to the partition sum, $Z(q,r)$, over the two distance scales 
produces each point in the plots above. The catalogue has approximately 1.5 million halo positions.}
\label{fig:Dq_curves}
\end{figure}

In \S\ref{2PCF} the 2PCF was related to the correlation dimension, $D_{2}$, 
through a differential operation. However, in Multifractal terms the 
correlation dimension is only one of an infinite number of generalised 
dimensions which we can use to characterise the distribution. Other important 
dimensions include, $D_{1}$ - {\em information} dimension, $D_{0}$ - {\em Capacity} 
dimension, and $D_{q}$ as $q\rightarrow\pm\infty$. Multifractal distributions are 
usually defined by a $D_{q}$ curve which generally decreases with $q$ (see figure 
\ref{fig:Dq_curves}). Monofractals on the other hand produce a flat $D_{q}$ curve.

\begin{figure}
\centering
\includegraphics[height=7cm,width=6cm,angle=-90]{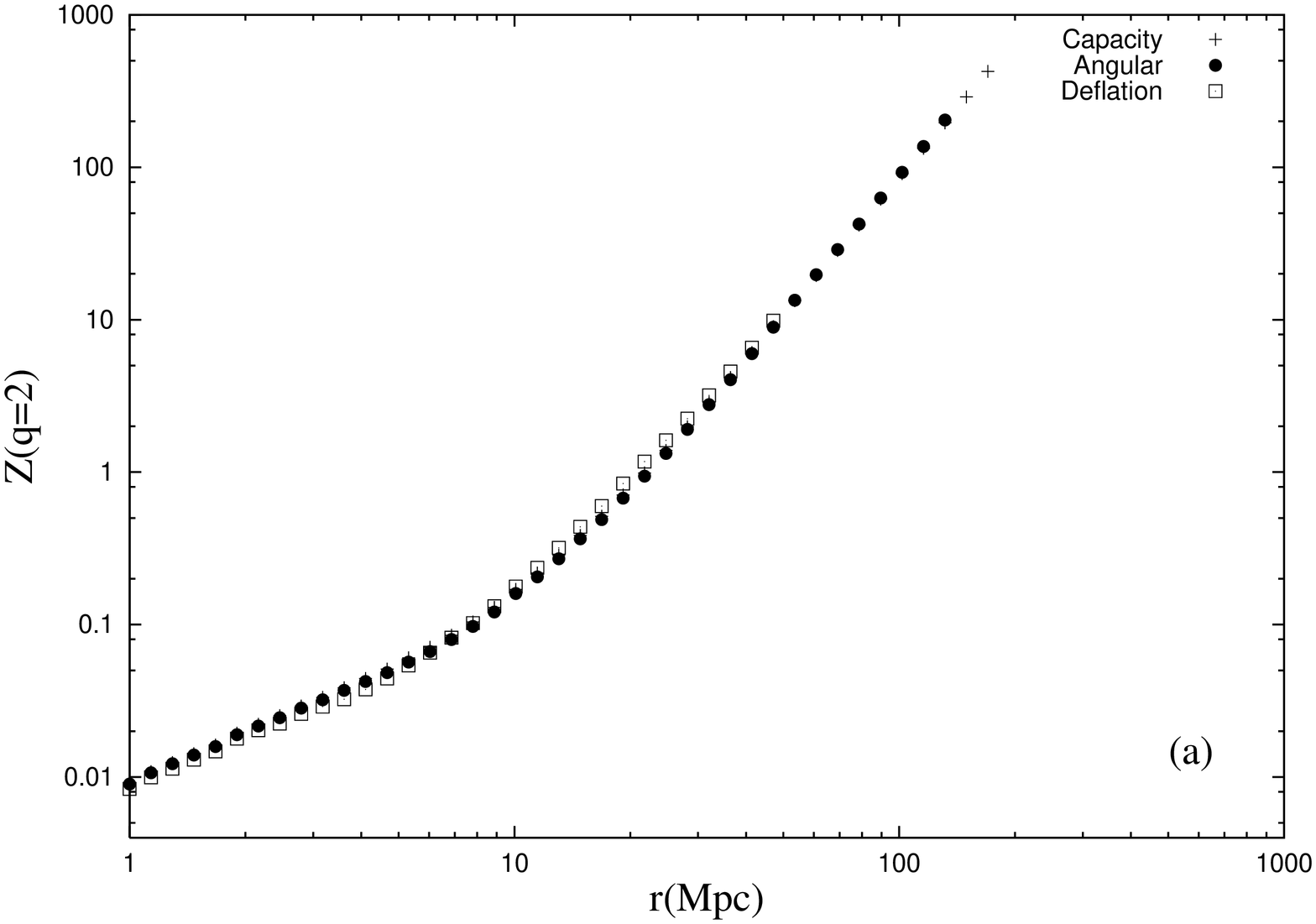}
\includegraphics[height=7cm,width=6cm,angle=-90]{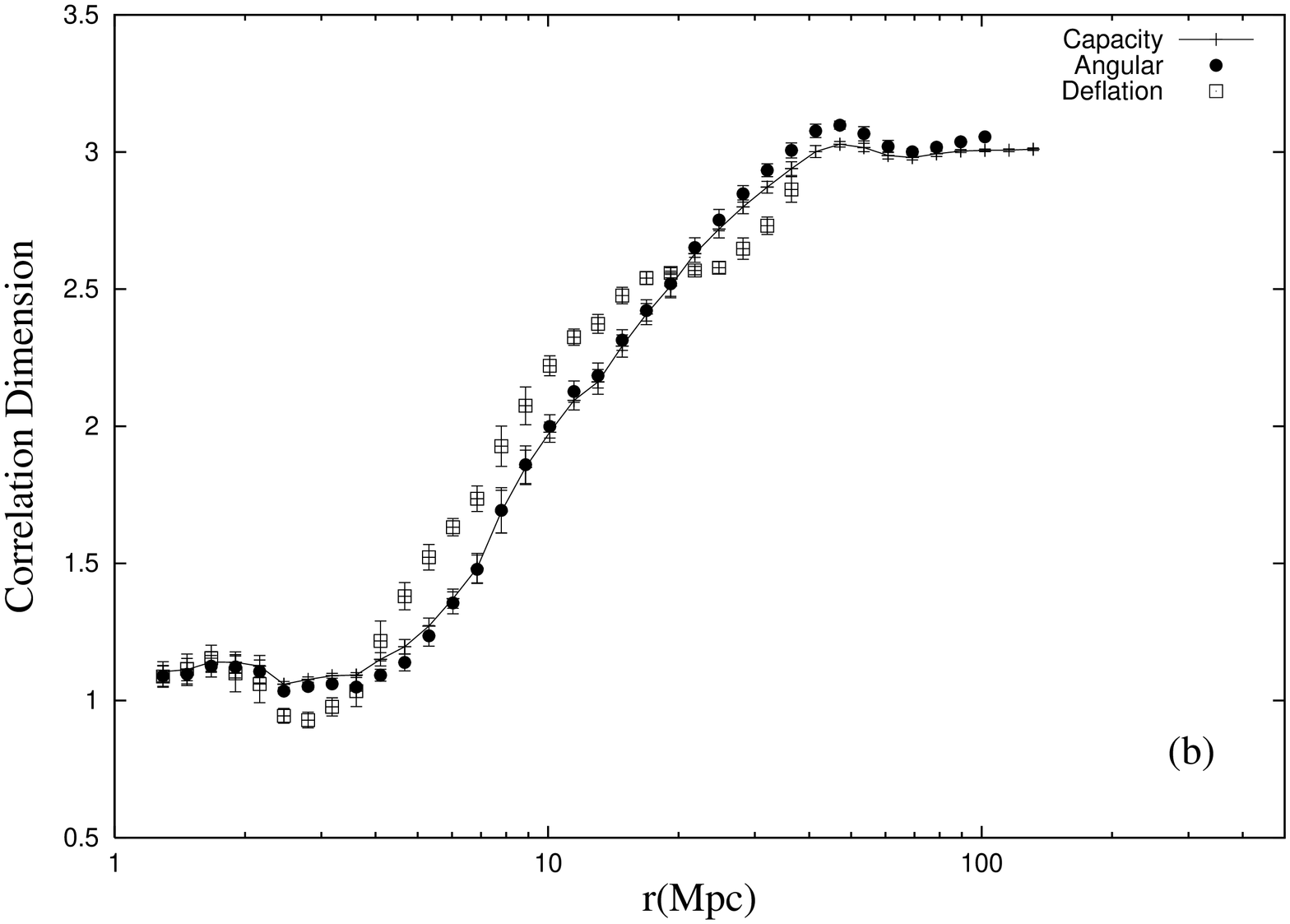}
\caption[The Partition sum and Correlation Dimension]
{\small Results of the mock \pscz~catalogue from a 
$\Lambda$CDM model. {\em Left:} The partition sum $Z_{2}$ varying with distance. This is very 
closely related to the 2PCF. The direct derivative of this plot can determine the correlation 
dimension varying with scale. $[$Figures from Pan \& Coles, 2001$]$}
\label{fig:partition_sum}
\end{figure}

The methodology we have constructed is used to calculate the $D_{q}$ curve for the 
$\Lambda$CDM $N$-body simulations. The two plots in figure \ref{fig:Dq_curves} show $D_{q}$ curves for 
different distance scales. A minimum $\chi^{2}$ approach was applied to the partition sum, $Z(q,r)$ 
as illustrated in figure \ref{fig:partition_sum}(a). 

The $\chi^{2}$ minimisation was performed on a straight line model fit to the data at two different 
distance scales; $10 < R_{1} < 40$ Mpc and $50 < R_{2} < 100$ Mpc. Both $R_{1}$ and $R_{2}$ were chosen as they 
appeared to have constant gradient in these regions. At small scales the data seems to be supporting 
a multifractal distribution, whereas on large scales the universe appears to reach homogeneity on 
scales considerably smaller than the size of the box.  

\subsection{Other Estimators}
The partition sum can also be calculated using a counts-in-cells approach (Mandelbrot, 1982).
Define a new measure $\mu$ of a discrete point set as,
\begin{equation}
\mu_{i}=\frac{N_{i}}{N_{tot}},
\end{equation}
which is just the fraction of all galaxies $N_{tot}$ contained within cell $i$. Also if the 
relation $\sum_{i}\mu_{i} = 1$  is satisfied, then it must follow that the cells cover the 
entire space (of topological dimension, $d$) and that the cubes have volume $r^{d}$. 
We can now construct a new measure,
\begin{equation}
M(q,r)=\sum^{N(r)}_{i=1}\mu_{i}^{q}.r^{a}=N(q,r).r^{a}\begin{array}{c} \longrightarrow \\ 
[-4mm] {r \rightarrow 0} \end{array}\left\{\begin{array}{cc} 0,& a > \tau(q),\\ \infty,& a < \tau(q),
\end{array}\right., 
\label{eq:box_counting}
\end{equation}
with $N(r)$ being the number of occupied cells and $N(q,r)$ is the number of occupied cells, 
weighted by $q$. The measure is dominated, for large positive values of $q$ when the cells are 
more populated, and for large negative $q$ when the cells are sparsely occupied. This is a very 
important property of multifractal analysis, since under-dense and over-dense regions are probed 
by different values of $q$.

In the limit $a\rightarrow\tau(q)$, it follows that $M$ will be finite and nonzero. Then we can see 
from eq.(\ref{eq:box_counting}) that,
\begin{equation}
N(q,r) = \sum^{N(r)}_{i=1}\mu_{i}^{q} \propto r^{-\tau(q)},
\end{equation}
and
\begin{equation}
\tau(q)=\lim_{r \rightarrow 0}\frac{\ln N(q,r)}{\ln (1/r)}.
\end{equation}

\subsection{The $f(\alpha)$ Curve}
Grassberger \etal~(1988) show that we can rewrite the usual fractal measure as,
\begin{equation}
p_{i}\sim r^{\alpha_{i}}
\end{equation}
The distribution of the scaling indices $\alpha_{i}$ characterise the 
dimensionality of the survey. This is evaluated using the $\alpha$-spectrum,
\begin{equation}
n(\alpha)d\alpha \sim N\vert\ln{r}\vert^{1/2}r^{\alpha-f(\alpha)d\alpha},
\end{equation}
where $n(\alpha)d\alpha$ is the number of times that $\alpha$ takes values in the range 
$(\alpha, \alpha+d\alpha)$. For a homogeneous fractal distribution the $f(\alpha)$ curve 
reduces to a single point: $\alpha_{0} = f(\alpha_{0}) = D_{0}.$ In any case the statistical 
properties of a distribution are equally described by either the generalised dimensions, $D_{q}$ 
or by the $f(\alpha)$ curve since they are a Legendre pair, as shown below. The only main 
drawback, as we will soon see, is that the latter strategy requires an extra differentiation 
of the data.

\begin{figure}
\includegraphics[scale=.5]{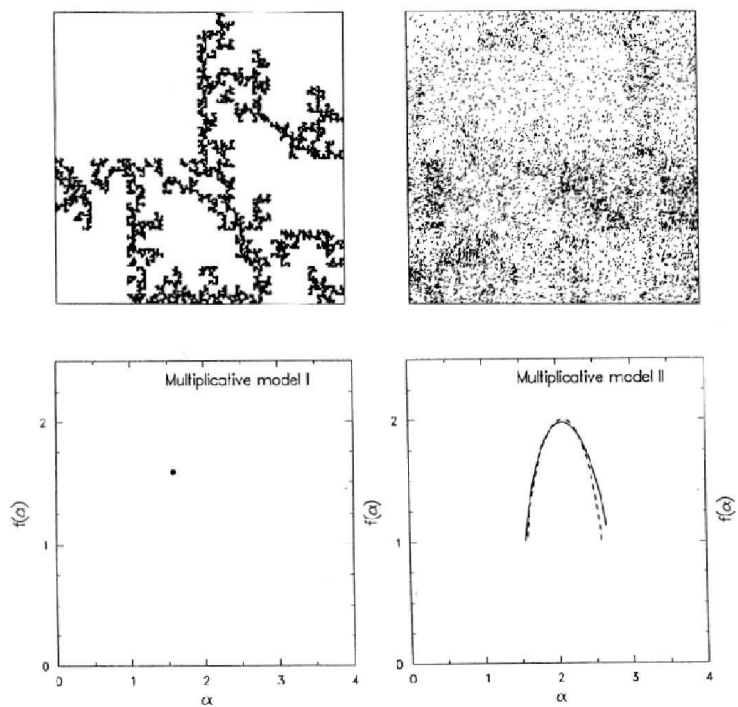}
\caption[Multifractal Model distributions]
{Two fractal distributions with their corresponding $f(\alpha)$ curves below. 
{\em left:} A monofractal distribution produces a single point in the $f(\alpha)-\alpha$ 
plane. {\em right:} A genuine multifractal point set produces a curve in the $f(\alpha)-\alpha$ 
plane. $[$Figures  from Jones \etal~1988$]$}
\label{multifractals}
\end{figure}

We can show this by considering the integral version of equation (\ref{partition}),
\begin{eqnarray}
Z(q,r)=\frac{1}{N}\sum^{N}_{i=1}p_{i}(r)^{q-1}
=\frac{1}{N}\int n(\alpha)r^{(q-1)\alpha}d\alpha, \nonumber \\
=\int\vert \ln r \vert^{1/2}r^{\alpha q -f(\alpha)}d\alpha.
\end{eqnarray}
A solution to the above expression can be found using the Laplace integral approximation 
(Martinez \etal~1990), giving,
\begin{equation}
Z(q,r)=r^{\alpha(q)-f\{\alpha(q)\}}\left\{\frac{\pi}{2f^{''}[\alpha(q)]}\right\}^{1/2}.
\label{eq:Z_approx}
\end{equation}
The conditions of this theorem defining the function $\alpha(q)$ are,
\begin{equation}
\frac{d \ f(\alpha^{'})}{d\alpha^{'}}\vert_{\alpha^{'}=\alpha(q)},
\label{condition1}
\end{equation}
and
\begin{equation}
\frac{d^{2}f(\alpha^{'})}{d\alpha^{'}} < 0.
\label{legandre}
\end{equation}
Using $Z(q,r)=const \times r^{\tau(q)}$ and (\ref{eq:Z_approx}) we get,
\begin{equation}
\tau(q)=\alpha(q)q-f(\alpha),
\label{tau_relation}
\end{equation}
and using (\ref{condition1}) with (\ref{legandre}) leads to,
\begin{equation}
\alpha(q)=\frac{d\tau}{dq}.
\label{alpha_relation}
\end{equation}
Equations (\ref{tau_relation}) and (\ref{alpha_relation}) relate the variable pairs, 
$(q,\tau)$ and $(\alpha,f)$: a Legendre transform. So we can see that the 
distribution is equivalently characterised using either method. However, in practical 
terms the Generalised Dimension approach may prove to be more accurate, since the 
$f(\alpha)$ curves require a further differentiation of the data. i.e. $f(\alpha)$ 
curves (fig.\ref{multifractals}) are related to the derivative of the $D_{q}$'s 
(fig.\ref{fig:Dq_curves}), through eq.(\ref{alpha_relation})

In figure \ref{multifractals}, two realisations of a multiplicative random fractal are 
shown with their corresponding $f(\alpha)$ curves. The method for constructing these distributions 
is discussed in Jones \etal~(1988).

\chapter{Boundary Corrections}
\vspace{-0.6cm}
As we touched on in \S\ref{problems}, the analysis of redshift surveys is prone to many
problems, some subtle and some not so subtle.  The presence of large `holes' in
a survey -- regions where no galaxies have been observed -- is clearly an
example of a serious problem which needs to be appropriately corrected for.

\begin{figure}
\begin{center}
\includegraphics[scale=.45,angle=0.0]{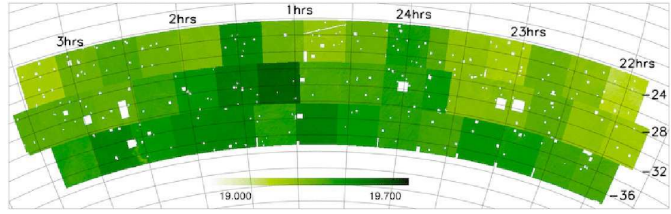}
\caption[The 2dFGRS mask locations and magnitude limits]
	{\small 
The 2dFGRS southern region shown in projection.  The colour scale gives the 
varying magnitude limit from plate to plate, white denotes regions which lie 
outside the survey. $[$Figure from the website: 2dF www.mso.anu.edu.au/2dFGRS$]$}
\label{fig:masks}
\end{center}
\end{figure}

To analyse real (or indeed mock) galaxy surveys we must, therefore, deal with the 
practical issue of incomplete sky coverage. This can arise firstly because of the 
geometry of the survey, which is usually a thin beam or a fan. Figure \ref{fig:masks} illustrates 
the latter case for the recent example of the 2dF galaxy redshift survey (2dFGRS, Colless \etal~1999).  
The figure shows the projected distribution of galaxies on the
plane of the sky, from which we see that the coverage of the survey is not all-sky --
\ie~we do not sample galaxies over $4\pi$ steradians. Also, we notice that there
are small patches and strips within the geometrical area of the survey which were 
not sampled.

Secondly, incomplete sky coverage can be caused by the extinction of light through
parts of our own galaxy, or regions being obscured by local objects. Hence, for
example, even redshift surveys such as the IRAS PSCz (Saunders \etal~2000) which set out 
to be all-sky are missing galaxies within a `mask' close to the plane of the
Milky Way galaxy.

Thirdly, of course, as we already discussed in \S\ref{problems}, redshift surveys are affected
by radial incompleteness, which we describe in terms of a selection function,
caused by the flux limit below which distance galaxies are too faint to be observed.
We can see the influence of this flux limit in Figure \ref{fig:masks}, which shows the variation
in magnitude limit from plate to plate in the 2dFGRS.

So, in summary, information about the `true' population of galaxies is hidden or
distorted by the effect of a flux limit, by the presence of masked regions and by
the boundary of the survey itself.

There has been a number of methods created to account for the problems mentioned above. In the following 
sections a few of these methods will be reviewed and in \S\ref{volcorr} a new correction technique is 
introduced, whose benefits include computational and statistical efficiency.

\section{Deflation Method}
\begin{figure}
\centering
\includegraphics[scale=0.5]{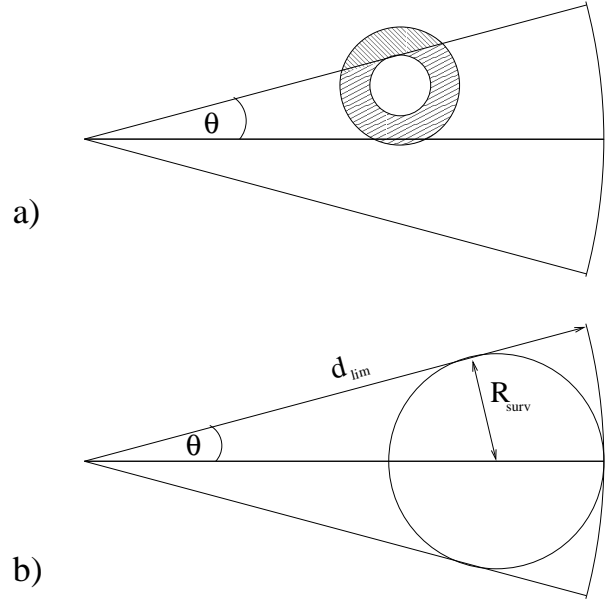}
\caption[Illustration of deflation method]
	{\small Panel (a) illustrates the state of concentric shells around 
a galaxy in the survey.  As the radius of the shell increases, eventually 
density is measured for shells that are partially external to the survey, 
such as the shaded shell above.  The lower panel (b) demonstrates the maximum 
distance to which this survey can be probed: the radius of the largest sphere 
which can totally be contained within the survey. [Figure from Hatton, 1999]}
\label{fig:deflation}
\end{figure}

As we consider placing spherical shells of increasing radius around a galaxy,
even allowing for our `weighting' of the number count of galaxies in each sphere
by the radial selection function of the survey, it is clear that we will eventually 
reach the edge of the survey (which of course we can think of as the distance
beyond which the selection function is equal to zero). Therefore, unless some form of 
edge correction is applied, further increases in the shell radius will result in an 
estimate of the density for shells that are systematically underpopulated relative to 
the mean density of the underlying galaxy population, because of the volume of each
shell that lies outside the survey.  This effect is illustrated in 
the upper panel of Figure \ref{fig:deflation}.  For the outer (shaded) shell in this diagram, the 
estimated density will be systematically lower than the true density since the shell 
includes a region that lies entirely outside the survey volume and so, by definition, 
will contain no galaxies.

The {\em deflation\/} method is, perhaps, the simplest and crudest form of boundary 
correction. It simply restricts the sum in equation (\ref{prob}) to include only those
counting spheres which lie completely within the survey region. However, this
drastically reduces the distance out to which the density estimator can reliably
probe, leading effectively to a `cosmic variance'\footnote{Cosmic variance 
is the cosmologist's definition of sample variance. If the scales being 
sampled are comparable in size to the survey, then only a few independent 
measurements can be taken, \ie~only a few values available for averaging.} 
problem at larger radii.

The maximum scale, $R_{surv}$ which is probed using 
the {\em deflation} method is defined as (Hatton, 1999),
\begin{equation}
R_{surv}=\frac{d \sin \theta_{surv}}{1 + \sin \theta_{surv}},
\end{equation}
where $\theta_{surv}$ is the opening angle of the survey and $d$ the 
volume limit.

As an example, the Stromlo-APM redshift survey (Loveday \etal~1992) 
had an opening angle, $\theta_{surv} = 22.6^{o}$ and a volume limit, 
$d \sim 110h^{-1}$Mpc, this leads to a maximum sampling radius of only, 
$R_{surv} = 30.5h^{-1}$Mpc. 

\section{Capacity Correction} 
The {\em capacity} correction can be thought of as the next step up from the {\em deflation} method. Here we allow 
all counting spheres while using equation (\ref{prob}), even those which cross the boundary or 
contain masks. The missing volume is accounted for by re-weighting the contribution of the 
(incomplete) counting sphere, essentially equivalent to filling it with a distribution of mock 
galaxies. The main problem is in deciding which distribution should be used to fill the void. 
Borgani \etal~(1994) chose to weight each cell by a factor $f_{i}(r)$ which is determined by 
the missing volume. 

This at first glance may seem to be a valid choice.  On the other hand the
weighting factor should be, more correctly, proportional to some measure of the 
average density in the counting sphere.  This highlights the potential problem with
the capacity correction: even though it can, in principle, be applied to considerably
larger counting spheres than the deflation correction, its form is fundamentally
flawed since it is assuming an answer to the question being posed -- \ie~that
$N \sim r^3$, which is a statement of exactly the homogeneity that we are trying to test.

\begin{figure}
\begin{center}
\includegraphics[scale=.8,angle=0.0]{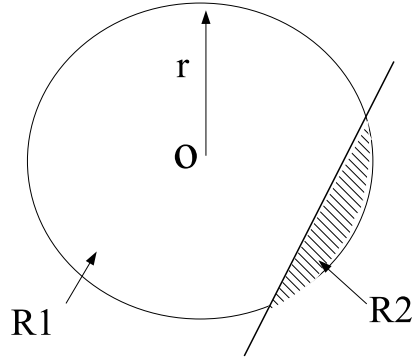}
\caption[Counting sphere illustrating boundary \& masked regions]{\small A counting sphere 
centred on galaxy O with radius r. Regions R1 \& R2 are inside and outside the survey 
respectively.}
\label{fig:circle1}
\end{center} 
\end{figure}

In practise the capacity correction is applied using the expression,
\be
p_{i}(r) = \frac{n_{i}(r)}{N_{tot}} = \frac{n_{i}^{\ast}(r)+\bar{\rho}.V_{missing}}{N_{tot}},
\label{eq:capacity1}
\ee
\be
=\frac{1}{N_{tot}}\left \{\displaystyle\sum^{N}_{j=1}\frac{\Phi(|r_{j}-r_{i}|-r)}
{\phi(r_{j})}+4\pi.\bar{\rho}(r^{2}.{\Delta}r+r.{\Delta}r^{2}+{\Delta}r^{3}).\frac{M}{M_{tot}}\right \},
\label{eq:capacity2}
\ee
to obtain the corrected number of objects within radius $r$ from a given particle. 
The RHS of eq.(\ref{eq:capacity1}) contains two main terms: the reduced number 
count, $n_{i}$, and a term accounting for galaxies which are missing. This is done 
by calculating the missing volume of the survey and filling it with a density 
corresponding to the average density of the survey. 

To determine the missing volume of the sphere one can place random points within
it, or equivalently one can shoot off vectors in random directions from
the central point in the sphere, and count how many of these lie within the survey.
This approach is known as a {\em Monte Carlo\/} technique. The expression for the 
missing volume, therefore, requires us to determine two numbers:
\begin{enumerate}
\item{the number $M$ of random vectors emanating from the central galaxy which
fall outside the survey}
\item{the total number $M_{tot}$ of random vectors emanating from the central galaxy.}
\end{enumerate}

Of course choosing to fill the missing portion of the counting cell with any 
presupposed density should be considered {\em bad science}. What we would ultimately 
like to do is fill the missing regions with the right amount of particles. To do 
this we need information which is hidden from us, behind the mask, but the 
Cosmological Principle may come to our rescue.

\section{Angular Correction} 
In 2002 Pan \& Coles, used the the assumption of isotropy\footnote{The Isotropy 
of the universe is a cornerstone of modern cosmology so in this case it is not a 
bad assumption to make.} to infer properties of unknown regions, masks and boundaries, 
from the well known survey space. Essentially they proposed that you can average over 
the part of the sphere which is observed and use this average to fill in for the 
missing regions. Of course this has to be done in an angular fashion, since $\xi(r)$ 
is assumed not to depend on direction.

\begin{figure}
\begin{center}
\includegraphics[scale=.8,angle=0.0]{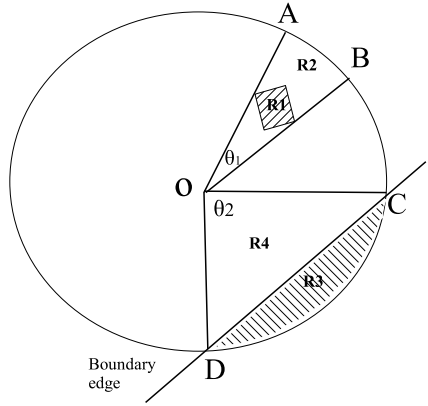}
\caption[Counting sphere illustrating the Angular correction]{\small This is a counting 
cell within the survey, it has a masked region (R1) and a missing portion due to the 
intersection with the boundary (R3). The slices AOB and COD encompass both of these 
missing parts.}
\label{fig:angular}
\end{center}
\end{figure}

Assuming that the universe is statistically the same in all directions, they concluded 
that the number of galaxies in a given solid angle should be comparable to the number of 
galaxies in the same solid angle but in a different direction. The average density of 
galaxies per steradian can then be defined. In order to achieve this, they introduce the 
weighting factor $f_{i}(r)$ which has values $0 < f_{i} < 1$. The value of the weight 
determines how much of the counting sphere is missing from the survey, and therefore 
needs to be accounted for. Then the corrected number of galaxies in a cell is computed from,
\begin{equation}
n_{i}(r)=\frac{1}{f_{i}(r)}\sum_{j=1}^{N}\frac{\Phi(|r_{j}-r_{i}|-r)}{\phi(r_{j})},
\label{eq:PC01}
\end{equation}
where $\phi$ is the selection function for a flux limited sample. The weighting factor 
appears in the above expression to increase the number count accordingly, just as the 
selection function does.

To implement the angular correction one would start as usual and centre on a galaxy within the 
survey. At a given radius the counting sphere may contain a part of a masked region 
(indicated by region R1 in fig.\ref{fig:angular}) or the survey boundary (R3 in fig.\ref{fig:angular}). If 
this were to occur, the solid angles which contains these features are cut out (slice AOB \& COD in 
fig.\ref{fig:angular}) and replaced by an average over the rest of the cell. As an example, for figure \ref{fig:angular} 
the weighting factor in this 2-D analogy is $f_{0}=1-(\theta_1 + \theta_2)/2\pi$. 
For computational purposes, the method to calculate equation (\ref{eq:PC01}) is as follows: 
\end{spacing}\begin{spacing}{1.0}
\newcounter{newcount}
\begin{list}{\small\arabic{newcount}}
{\usecounter{newcount}
\setlength{\rightmargin}{\leftmargin}}
\item From the set of all galaxies, choose a galaxy as the centre of a counting sphere
of radius $R$ and determine its position relative to the mask and boundary.
\item `Shoot off' random vectors ($r,\theta,\phi$) from the centre, where:
	\begin{itemize}
	\item{ } $r$ is sampled uniformly $\in (0,R)$. $R$ - sphere radius.
	\item{ } $\phi$ is sampled uniformly $\in (0,2\pi)$.
	\item{ } $\theta=\sin^{-1}(U)$, U is sampled uniformly from $\in(-1,1)$.
	\end{itemize}
\item Determine which vectors lie outside the survey region. Any vector which does so, save its components.
\item Using the scalar product expressed in terms of vector components, calculate
      the angle between each pair of vectors which lie inside the survey. This
      allows us to define a reference direction, and a maximum angle, $\theta_{max}$ 
      between vectors which lie inside the survey. ($\theta_{max}$ is the opening
      angle of the cone which points in the reference direction).

\item Now begin counting those galaxies in the counting sphere which make an angle, $\theta > \theta_{max}$ 
      relative to our reference direction.
\item The weighting factor, $f_{i}(r)$, is then simply related to $\theta_{max}$.
\end{list}
\end{spacing}
\begin{spacing}{1.5}
This is essentially the algorithm developed by Pan \& Coles (2002) which they
tested on simulated catalogues and applied to analyse the fractal clustering of
the IRAS PSCz survey.

The angular correction, although in principle a very successful method to
correct for boundary effects, is very slow and inefficient.  Moreover, and worse
still, it throws away potentially useful data.  We can see this from Figure 4.4,
where the galaxies in regions R2 and R4 are excluded by the angular correction.

In the next section we consider a method which has the potential to improve further
upon the angular correction.

\section{Volume Correction}  
\begin{figure}
\begin{center}
\includegraphics[scale=.8]{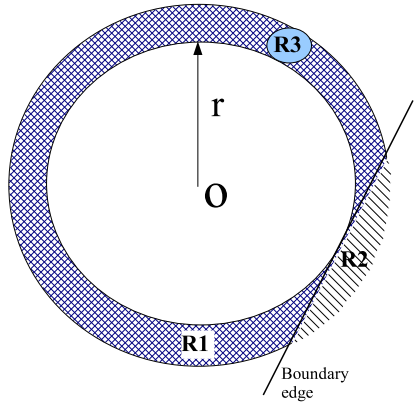}
\caption[Counting sphere illustrating the Volume correction]
{\small A counting shell centred on galaxy O with radius r. Region R1 is inside the survey, 
R2 \& R3 are outside the survey. The missing parts of the shell (R2 \& R3) are replaced by the 
average over the rest of the shell.}
\label{fig:circle}
\end{center}
\end{figure}

The basic idea of our new, volume, correction can be illustrated in figure \ref{fig:circle1}. 
Here the counting sphere has exceeded the geometrical 
boundary of the survey. The number of galaxies counted in the sphere of radius 
$r$ is depleted which leads to $p_{i}(r)$ being reduced through equation 
(\ref{prob}). As we have seen, to solve this problem we could either add galaxies to the 
missing region, as is the case with the {\em capacity} correction, or or equivalently we 
could somehow modify our definition of the volume itself (hence the name for our new
correction!). Of course  you may notice that eq.(\ref{prob}) does not contain any explicit reference to the 
volume, but we can cast this equation as,
\begin{align}
{p}_{i}(r) & =\frac{{V}_{i}(r)\rho^{\star}(r)}{{N_{tot}}}, \\ 
           & =  \frac{{V}_{i}(r)}{{V}_{i}^{\star}(r)}.\frac{n_{i}^{\star}(r)}{N_{tot}},
\label{volcorr}
\end{align}
with $V$ being the true volume of the sphere and $V^{\star}$ being what we can 
term the reduced volume. We have also introduced the reduced density, $\rho^{\star}(r)$, 
as an intermediary step which need not be calculated, and related this with a reduced 
volume and number count, $n_{i}^{\star}$. On its own this method can be visualised in 
figure (\ref{fig:circle1}), as assigning to the missing region (R2) the same density as that 
of region (R1). This would be wrong if density varies with distance, so that 
$\rho_{R1}\neq \rho_{R2}$. To overcome this problem we assume {\em only} that the 
density does not vary with $\theta$ or $\phi$ \ie~the universe is isotropic and 
hence equation (\ref{volcorr}) will hold for fixed $r$. So to apply this method to 
a galaxy survey we must count in spherical shells, correcting our estimate of the density in each shell as we
go along, and then integrate up the shells at the end. This method is illustrated in figure (\ref{fig:circle}). 
The shells are individually corrected and summed according to,
\begin{equation}
p_{i}(r)=\sum^{r}_{r=0}\alpha_{i}(r)\frac{n_{i}^{\star}}{N},
\end{equation}
where $\alpha_{i}(r)\equiv \frac{V}{V^{\star}}$; this is the
enhancement factor of the $i^{th}$ shell at radius $r$ and has value
$\geq$ 1.

The main advantage of this method is that makes the maximum use of the data. 
Specifically, if the boundary edge cuts across a counting sphere at a particular
radius, the method still makes full use of galaxies at smaller radii from the
centre of the counting sphere, even if they lie in the solid angle subtended by
the boundary edge.

The deflation method and the angular correction, of Pan \& Coles, on the other 
hand throw away a lot of potentially useful data, which limits the counting sphere 
radius within which the density may be reliably estimated.

\section{Practical Computing Issues}
One significant drawback faced when implementing the {\em angular} 
correction is that it must fill the counting spheres with random vectors to 
determine the missing regions. To obtain the required resolution means placing 
many random vectors at the centre of the sphere. For a cell with $R \approx 100\hm$ we have found that this requires the generation of $\sim 100,000$ random points.This is not a trivial computational 
task given that the vectors must be computed and angles stored at every 
step.

\begin{figure}
\begin{center}
\includegraphics[height=8cm, width=11cm]{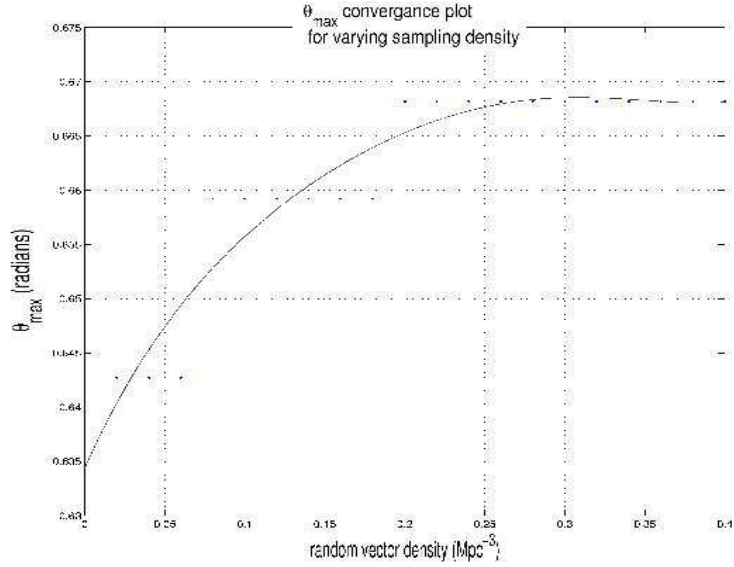}
\caption[$\theta$ convergence graph for Volume Correction]
	{\small The convergence of $\theta_{max}$ for increasing vector density. 
	The graph seems to converge at a random vector density of $\approx 0.25{\rm Mpc}^{-3}$. }
\label{fig:convergance}
\end{center}
\end{figure}

To see where this comes from consider figure \ref{fig:convergance}. Here a simulation has been 
set up to mimic what happens when 
computing the angular correction. A counting sphere is placed close to the survey boundary 
and $\theta_{max}$ is calculated for varying numbers of random vectors. From the plot we can 
see the convergence of the opening angle at an approximate vector density of 
$\rho_{vec}\sim0.25 h^{3}{\rm Mpc}^{-3}$. Taking this value we can make a back of the envelope calculation 
of the required number of random vectors to accurately constrain $\theta_{max}$ at a typical scale 
of $100 h^{-1}$Mpc. The number of vectors required is then,
\begin{eqnarray*}
N(r) &=& \rho_{vec}.V(r)\\
N(r) &=& \rho_{vec}.\frac{4\pi}{3}.r^{3}\\
N(r=100) &\approx& 100,000
\end{eqnarray*}
To make the comparison with the new volume correction, the same calculation is performed. 
In this case however we are calculating the number of vectors required to populate a spherical 
shell with thickness $dr=1$Mpc, as a typical value.
\begin{eqnarray*}
V(r)&\approx&4{\pi}r^{2}dr\\
N(r)&=&4{\pi}{\rho}r^{2}dr\\
N(r=100)&\approx&30,000
\end{eqnarray*}
The volume correction is definitely more computationally efficient, especially when you consider 
that the missing volume must be calculated for every galaxy and at every distance iteration. 
One should also bear in mind however that these codes could be used on much bigger surveys, 
SDSS is now pushing 1 million galaxies.

\section{Error Analysis}\label{errors} 
Error estimation in fractal analysis has been largely swept under 
the carpet by many in this field. One of the reasons for this is
the computational costs of using, \eg, a set of mock galaxy catalogues to estimate
error estimates via a Monte Carlo approach. Nevertheless, we can obtain an
approximate expression for the error on an estimate of the galaxy number density
via some remarkably simple mathematics.

We follow Grassberger and Procaccia's (1983) suggestion and use the partition function $Z(q, r)$, 
\begin{equation}
Z(q, r) = \lim_{r\to0} \lim_{N\to\infty} \sum _i ^M p_i^{q-1}  \sim  r^{\tau(q)} , 
\label{eq:partition2}
\end{equation}
to estimate the generalised dimensions of a set of  $N$ galaxies.  Here $p_i$ is the probability, 
\begin{equation}
p_i = \frac{n_i}{N},
\end{equation}
of a cell,  centred on the $i^{th}$ galaxy, having an occupation number $n_i$.  $r$  is the cell size while 
$\tau(q)$ is a scaling exponent. 
\begin{equation}
\langle Z(q) \rangle = \langle \sum_{i=1}^{M}  p_i^{q-1} \rangle
\end{equation}
By construction  the mean of the partition function is positive as there are always 
cells with galaxies within. However, there is a subtlety when handling astronomical data:
in expression (\ref{eq:partition2})  we assume that any set contains an infinite number of elements
and the size of the cells vanish. Redshift surveys, however, contain a finite number of objects 
in a finite volume. This will lead to configurations where none of the cells have galaxies other 
than the central one when $\lim {r\to0} $.

We can then construct the second moment of the distribution and rearrange as below,
\begin{eqnarray}
 &&\langle Z(q)  Z(q') \rangle  =  \langle  \sum_i  p_i ^{q-1}\,\sum_j p_j ^{q'-1}\rangle \\ 
 && =  \langle  \sum_{i,j}  p_i^{q-1} \,p_j ^{q'-1}\rangle \\ 
 && =  \sum_{i,j} \langle  p_i^{q-1}\,p_j^{q'-1}\rangle \\ 
 && =  \frac{1}{N^{q+q'-2} } \sum_{i,j} \langle  n_i^{q-1}  \, n_j^{q'-1}  \rangle	
\end{eqnarray}
\begin{equation}	
  =  \frac{1}{N^{q+q'-2} } \sum_{i=1}^{M} \langle  n_i^{q+q'-2}  \rangle + \frac{2}{N^{q+q'-2}}
 \sum_{i=1}^{M} \sum_{j = i + 1}^{M}  \langle  n_i ^{q-1} \,n_j ^{q'-1} \rangle.
\end{equation}
Where $\langle \cdots \rangle$ represents an ensemble average. Applying the {\it Cosmological Ergodic Theorem}
to the above equation leads to sums over the cells.
\begin{figure}
\begin{center}
\includegraphics[scale=0.42]{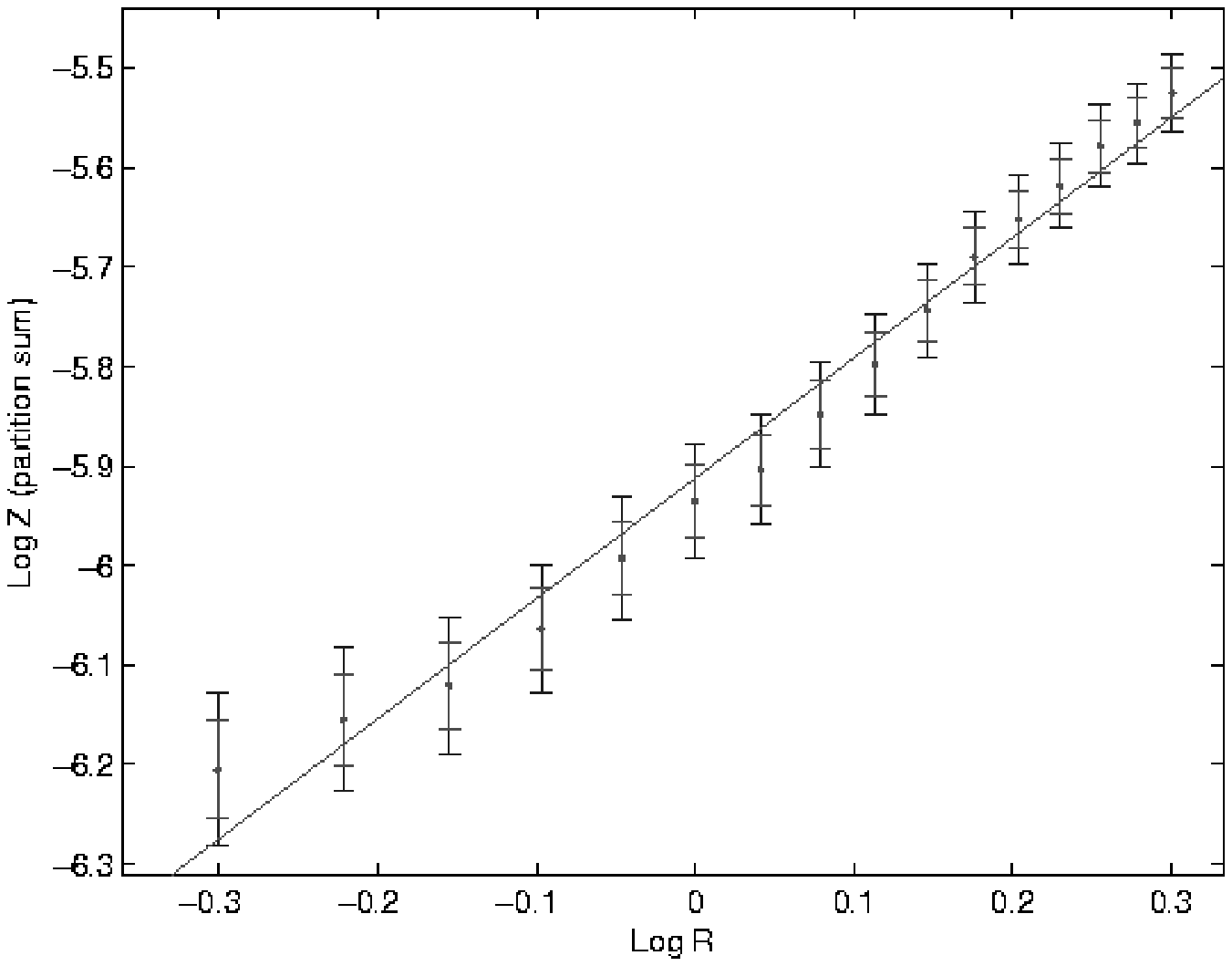}
\includegraphics[scale=0.42]{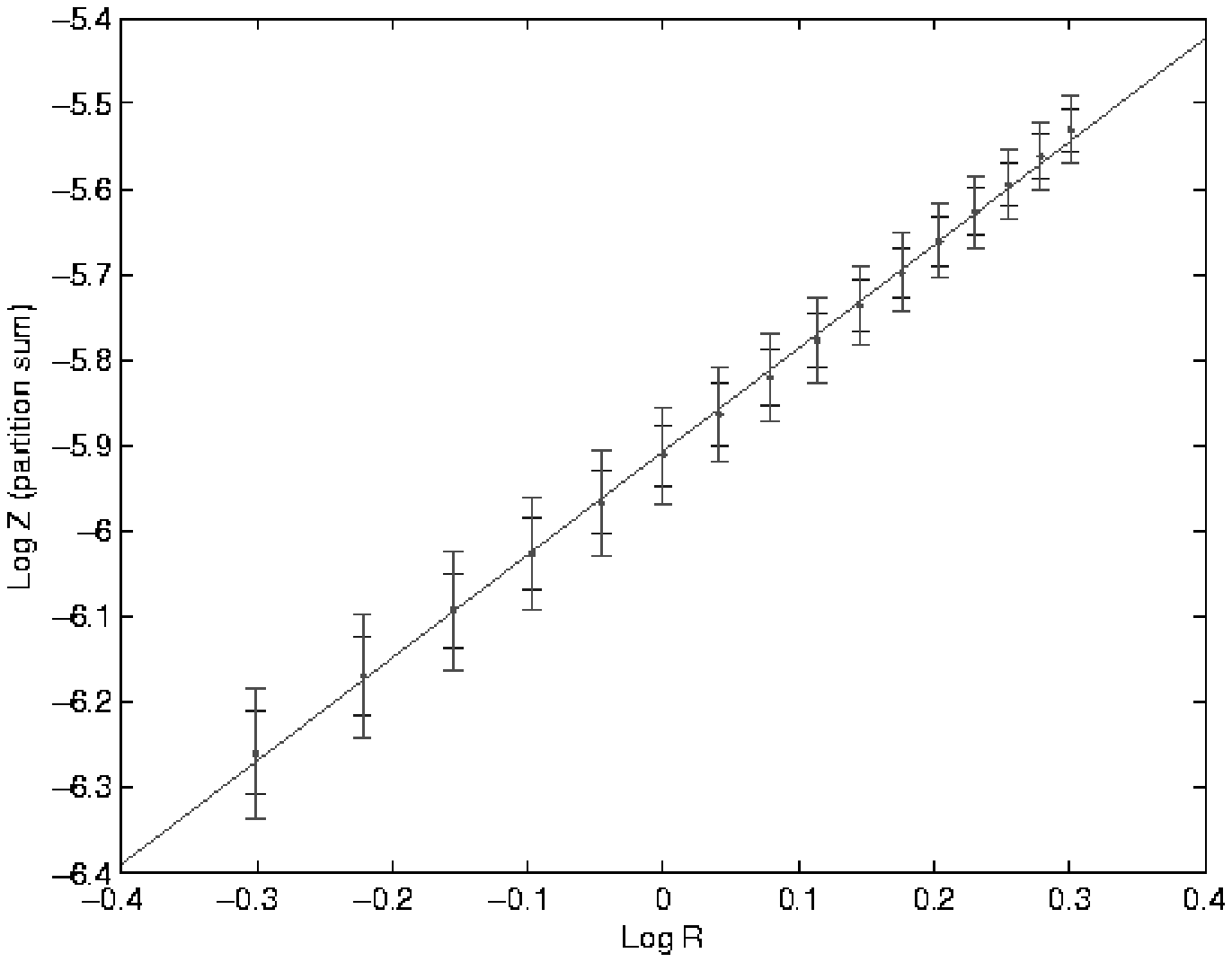}
\caption[Error comparison for multifractal analysis]
	{\small This is a log-log plot of the partition function varying with distance. 
	The larger  error bars are estimated from the prescription discussed 
	in \S~\ref{errors}, whereas the smaller  errors are obtained from averaging 
	the results of 100 L\'{e}vy Flight fractal simulations. This fractal has 1,000,000 particles 
	with $D_{2}=1.2$ and is contained in a $400Mpc^{3}$ box. {\em left:} The points in this plot 
        have not been averaged.  {\em right:} Points have been averaged over 100 distributions.}
\label{fig:fractal_errors}
\end{center}
\end{figure}
Thus, for $q=q'$ the $\langle Z(q)  Z(q') \rangle$ can be cast as 
\begin{eqnarray}
\langle Z(q)  Z(q) \rangle  & = &  \sum_i \langle  p_i ^{2(q-1)} \rangle  +  {\mbox{cross terms}}\\
 	& = &  \langle  Z(2q-1)   \rangle  +  {\mbox{cross terms}}\\
	& = &  \sum_i \langle  \left (p_i^2 \right ) ^{(q-1)} \rangle   +  {\mbox{cross terms}} \\
	& \le &  2\, \sum_i \langle  \left (p_i^2 \right ) ^{(q-1)} \rangle .
\end{eqnarray}
We can then estimate the standard deviation directly from,
\begin{equation}
\sigma_{q}^{2} \simeq {2.Z(2q-1)-Z(q)^2}
\label{eq:error}
\end{equation}
We show a comparison between our error estimation, from eq.(\ref{eq:error}), and 
an error estimation from 100 simulated L\'{e}vy Flight fractals (see figure 
\ref{fig:fractal_errors}).

The left hand plot, of figure (\ref{fig:fractal_errors}), shows the partition function  
calculated as a function of distance. To each point is attached a larger error bar 
estimated from the prescription discussed in \S\ref{errors} and a smaller error bar 
obtained via an average over the 100 simulations. We  can see from the plot that the 
gradient of $Z$ is constant and the points are all very close to the fitted straight 
line. The fit was obtained by minimising the $\chi^{2}$ function. The right hand plot 
shows similar points but now the partition function calculated at each distance is also 
averaged over the 100 simulations, with error bars computed as before. The excellent 
agreement in the fitted slopes of the left and right hand plots indicates that the partition 
function is an unbiased estimator. 

\chapter{Analysis \& Results}\label{applications}
\vspace{-0.6cm}
We will begin in the proceeding section by applying the different corrections 
to a toy fractal model. This model can be used to compare and contrast the 
differing methods. Then the {\em volume} correction will be used to analyse 
in detail the distribution of particles produced from an $N$-body simulation. 

\section{The L\'{e}vy Flight}
An initial test for the different corrections is to analyse a simple fractal distribution, 
the L\'{e}vy Flight. This fractal is very easy to construct and has an analytical 
determined dimension (see Meakin 1998).

\begin{figure}[h]
\centering
\includegraphics[width=5.3in]{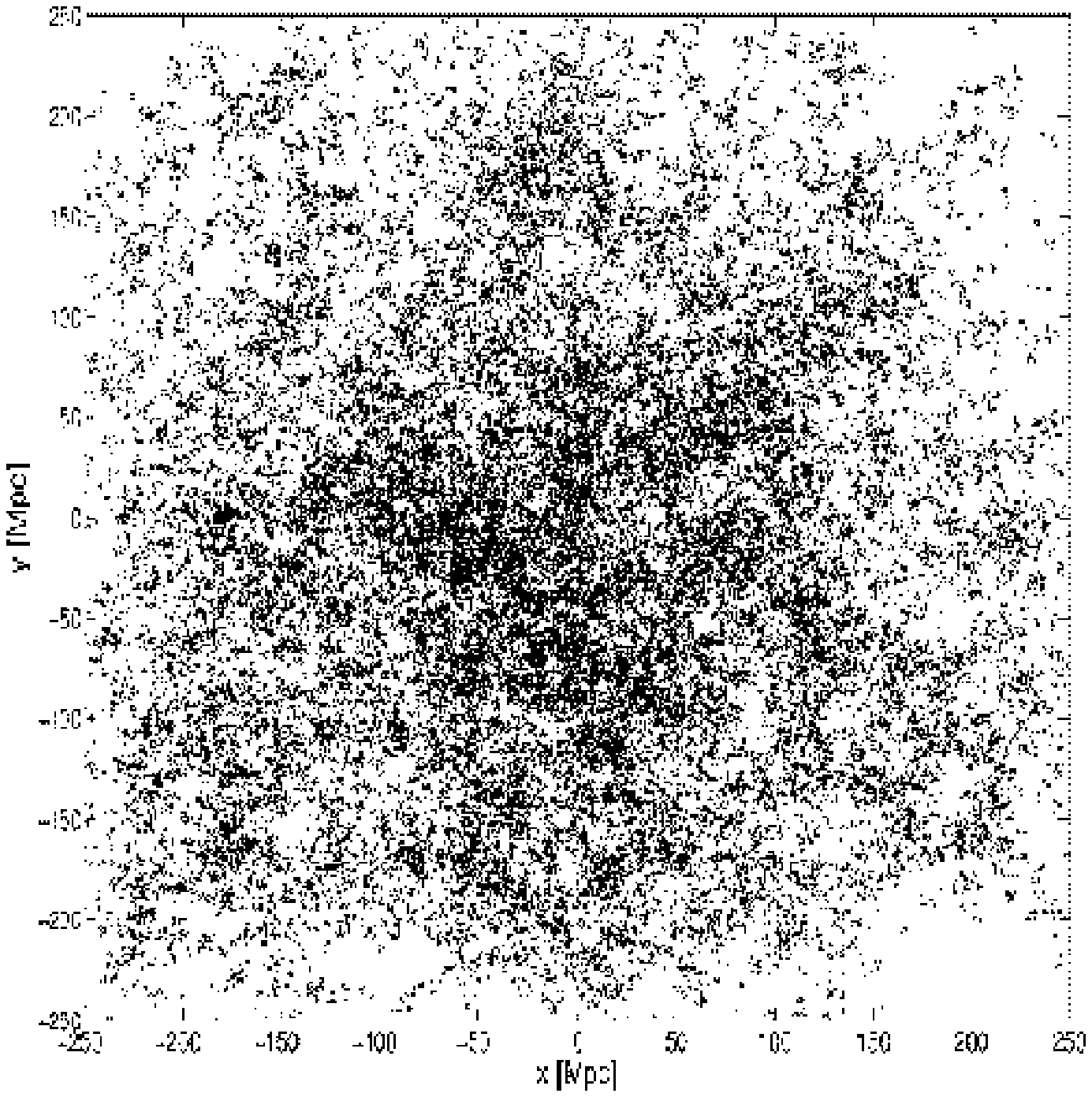}
\caption[L\'{e}vy Flight distribution of particles]
	{\small This is a 2-D view of a 3-D L\'{e}vy Flight distribution of particles,
	with parameters: $r_{0}=0.2$, $D=1.2$}
\label{fig:levy}
\end{figure}

The L\'{e}vy flight fractal is finding its way into many areas of physics due to 
its close connection to Brownian Random Motion. For example it has been used to 
explain Interstellar scintillation (Boldyrev \& Gwinn 2003) and even modelling 
the financial market (Chowdhury \& Stauffer 1999)\\
The L\'{e}vy Flight is constructed as follows:
\end{spacing}
\begin{spacing}{1.0}
\newcounter{LLLcount}
\begin{list}{\small\arabic{LLLcount}}
{\usecounter{LLLcount}
\setlength{\rightmargin}{\leftmargin}}
\item A point A is chosen at random, maybe the origin, in Cartesian space.
\item A displacement is given to A by a vector ($\theta$,$\phi$,$R$) to give a 
second point B. The angular direction is uniformly sampled and the Probability 
of $R$ exceeding a value r is given by,
\begin{equation}
P(R \geq \frac{r}{r_{0}}) = \left\{\begin{array}{cc} (r/r_0)^{-D},& r/r_0 > 1\\ 1,
& r/r_0 \leq 1
\end{array}\right.,
\label{eqa} 
\end{equation}
\item This procedure is repeated many times to `fill' the 3-D space, which was 
restricted to a cubic box of side 400 Mpc.
\end{list}
\end{spacing}
\begin{spacing}{1.5}
In expression (\ref{eqa}), $D$ is the fractal dimension and $r_0$ is a characteristic scale length, 
both of which we can adjusted to produce different features. The resulting distribution is not 
so dissimilar to a true galaxy survey (\cf~figure \ref{fig:levy}). $r_{0}$ can be related to 
the average inter-cluster separation.

\begin{figure}[h]
\centering
\includegraphics[height=3.2in,width=5.6in]{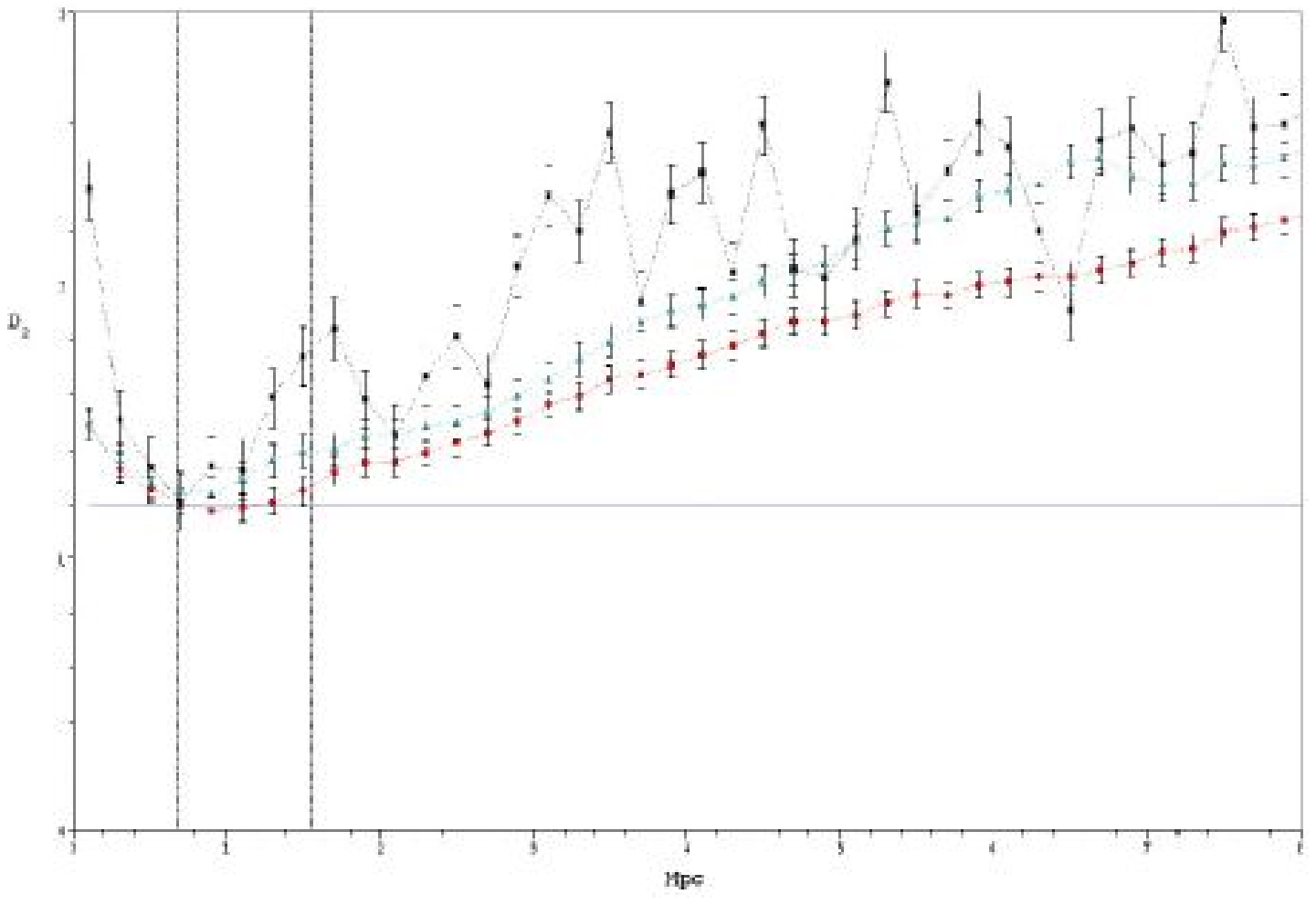}
\caption[$D_{2}$ evaluated for a L\'{e}vy Flight distribution]
	{\small $D_{2}$ evaluated for a L\'{e}vy Flight distribution of particles. 
	The volume (red), capacity (blue) and angular (black) corrections are used to 
	correct for the boundary of the simulation. The error bars are from our 
	prescription as described in \S\ref{errors}. The solid blue line corresponds to the 
	analytically determined $D_{2}$ value and the two vertical dashed lines represent the 
	average nearest neighbour separation and the mean inter-particle spacing respectively.}
\label{fig:D2_levy}
\end{figure}

\subsection{Multifractal Analysis}
A multifractal analysis, as described in \S~\ref{MFA}, is performed on a L\'{e}vy Flight distribution of 
particles (\cf~fig \ref{fig:levy}). The boundary is corrected by considering each of the different 
methods from Ch.4. This setup should give a fair comparison of the different correction methods. 

Since the correlation dimension is known analytically, the multifractal analysis will be restricted to 
the $D_{2}$ dimension. Now from the partition sum $Z(2,r)$, equation (\ref{eq:dimension}) will provide the $D_{2}$ 
value through differentiation. This differentiation was performed directly on the data by applying a 
linear fit to every three consecutive data points. The errors, as described in \S~\ref{errors}, were 
considered when minimising the $\chi^{2}$ function.

The results of this procedure are plotted in figure \ref{fig:D2_levy}. There are a few points to note 
in this plot. Firstly, the L\'{e}vy Flight is highly anisotropic, which is a cause for concern when 
applying a statistic which does not consider angular information. Although this may be a problem, 
following the suggestion of Mart\'{i}nez \etal~(1990), the correct result from the simulation 
seems to be confined to the range between the average nearest neighbour separation and the mean 
inter-particle spacing. In figure \ref{fig:D2_levy} it can be easily seen that the volume correction 
lies closer to the analytical answer of $D_{2}=1.2$, than either the capacity or angular corrections.

Secondly, the volume correction is more or less always closer to the true value, even as the methods 
begin to over estimate on larger scales. This over estimation by all the methods is due to the 
anisotropic nature of the L\'{e}vy Flight distribution. 

Thirdly, the large errors and visual noisiness of the angular correction are not present in either the 
capacity or volume corrections. This, I can only conclude, is due to the angular method throwing away 
data, leading to low number statistics, \ie~intrinsic noise.

\section{$\Lambda$CDM Simulation}\label{LCDM}
In this section we will analyse the distribution of dark matter halos from a $\Lambda$CDM simulation. 
Since there are no galaxies in this analysis, it is only the underlying dark matter distribution 
which is being probed. This simulation was performed by Warren \etal~(2006), see reference for a 
detailed description. For this work we are using a 384\hmpc~box with a flat geometry and cosmological 
parameters,
\be
\mathbf p = (\Omega_M, \Omega_b,n,h,\sigma_8) = (0.3, 0.04, 1, 0.7, 0.9).
\label{cosmoparams}
\ee
Initial conditions were derived from the transfer functions as calculated by CMBFAST 
(Seljak \& Zaldarriaga, 1996). The final catalogue has approximately 1.5 million halo positions. 

\subsection{Multifractal Analysis}\label{LCDM_MFA}
The multifractal analysis as presented in \S\ref{MFA}, must be applied over a range of different distance 
scales. In figure (\ref{fig:Dq_curves}) the results of this analysis are shown over two different 
ranges; $10<R_{1}<40 h^{-1}$Mpc and $50<R_{2}<100 h^{-1}$Mpc. These scales were chosen almost 
arbitrarily\footnote{The only reason was that the partition sum, $Z(q,R)$, seemed to 
have a constant gradient in these regions.}. The left hand plot of fig.(\ref{fig:Dq_curves}) 
shows a clear sign of multifractality on small scales whereas the right hand plot appears to signal 
a transition to homogeneity. However, since the partition function (like figure \ref{fig:partition_sum} 
(a)) is generally smooth, we can expect a smooth transition to homogeneity.

Instead of applying this analysis in certain ranges, the plots in figure \ref{fig:Dq_curves} could 
be extended by adding another axis: distance. This would give us a {\em $D_{q,R}$ Surface}.

\subsection*{The $D_{q,R}$ Surface}
\begin{figure}[h]
\centering
\includegraphics[scale=.6]{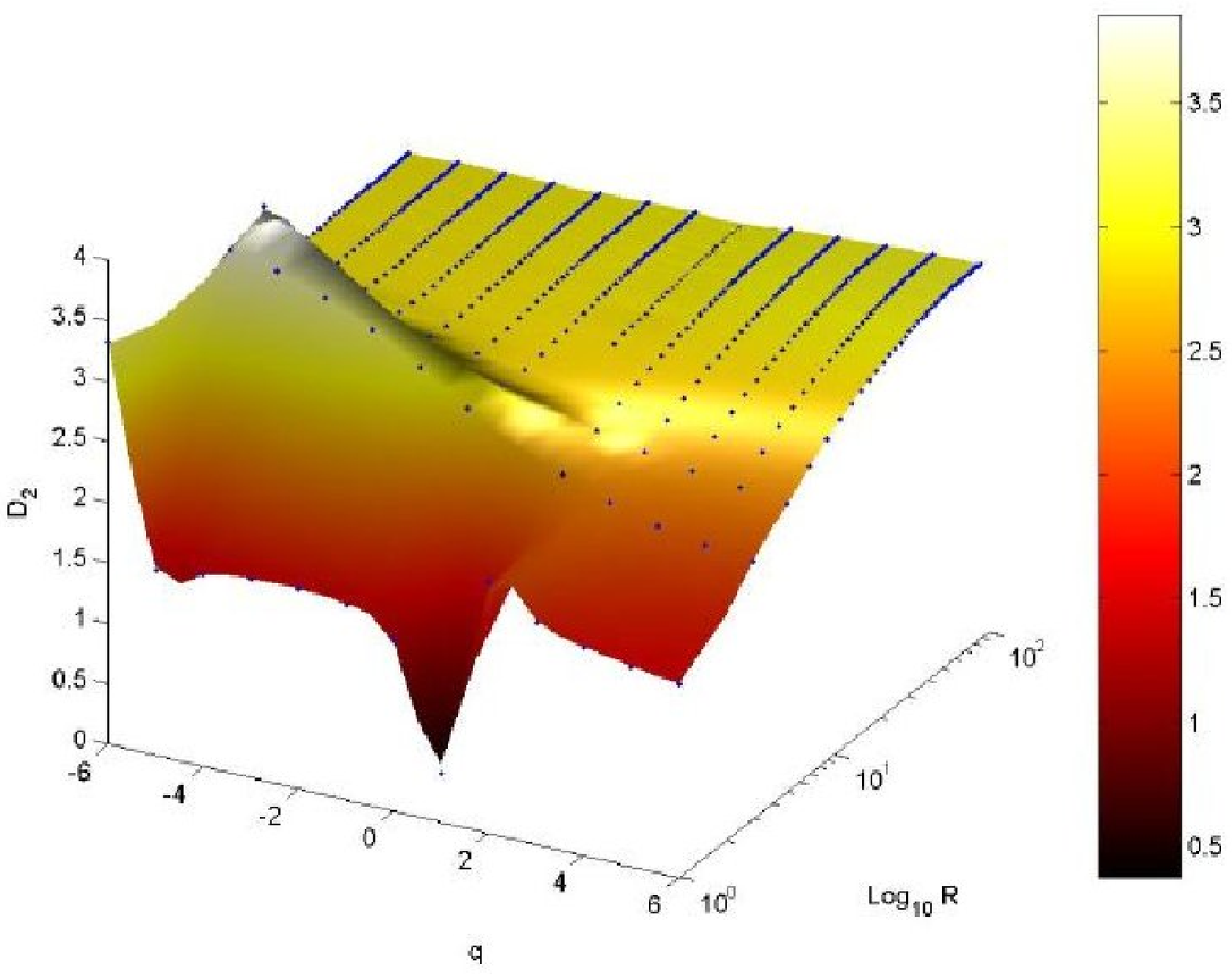}
\caption[$D_{q}$ varying with distance $\&$ moment, q.]
{\small This is a 3-d surface fit of $D_{q}(r)$. 
The data used is the halo positions from a $\Lambda$CDM cosmology. Described in Warren \etal~2006.}
\label{fig:3d-fit}
\end{figure}

\begin{figure}
\centering
\includegraphics[scale=.8]{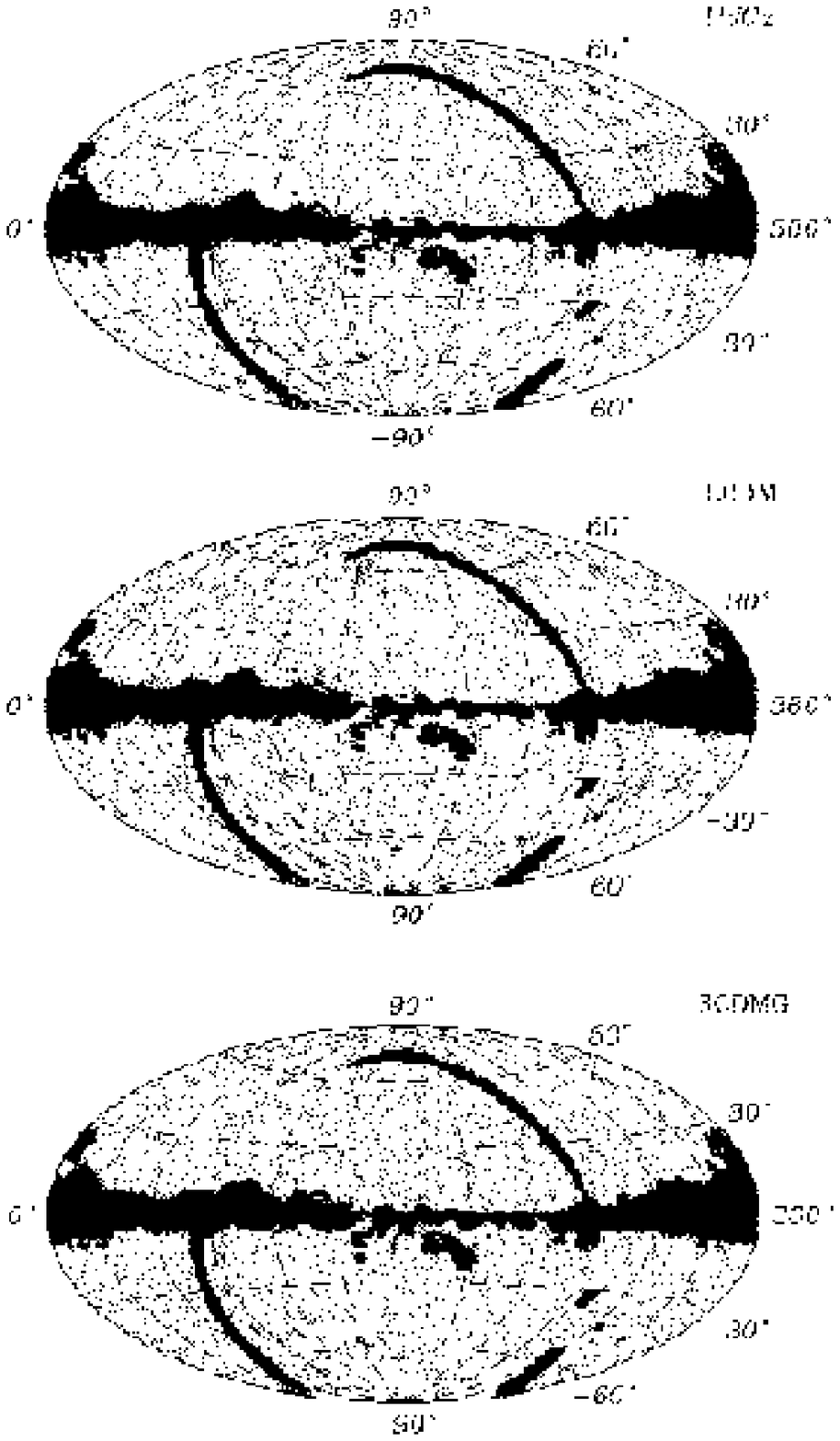}
\caption[Aitoff projection of PSCz and mock catalogues]
	{\small 
Sky distribution of galaxies in the \pscz~and in two $N$-body mock-catalogues. 
From the top to the bottom we illustrate mocks drawn from the LCDM and SCDMG 
cosmologies, respectively. The Aitoff projection is in Galactic coordinates.}
\label{aitoff}
\end{figure}

\begin{figure}
\centering
\includegraphics[width=5in]{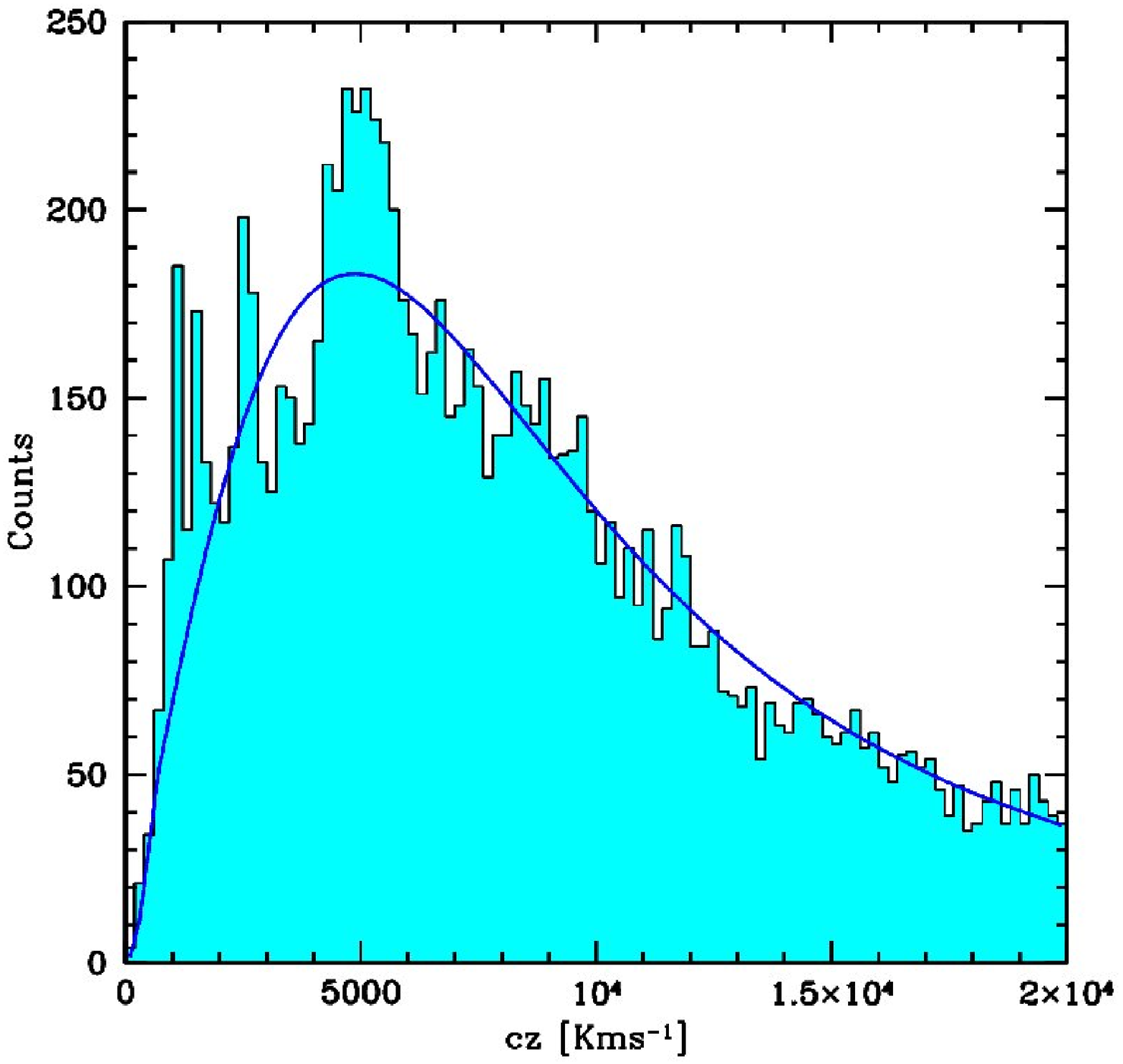}
\caption[Histogram of the PSCz survey]
	{\small Histogram of the radial number count, $dN$, in the PSCz survey. 
	The solid line represents the expected number in each bin after the inclusion of the selection function,$\phi$. 
	\ie~$dN = 4{\pi}\bar{n}\phi(r)r^{2}dr$.}
\label{fig:histogram}
\end{figure}

Figure \ref{fig:3d-fit} shows the result of extending the multifractal analysis to include
varying scales. The high peak and dip on low scales ($<10 h^{-1}$Mpc) corresponds to a multifractal
distribution. The surface then levels off to a constant value of three advocating a transition
to homogeneity at a scale of $\approx~30 h^{-1}$Mpc. Another interesting feature is that homogeneity is 
not reached at the same scale for all $q$.

\section{PSCz Mock Catalogue}
In the previous section the whole halo catalogue was used, thus the analysis 
is only relevant for the underlying dark matter distribution. In this section 
the galaxies are under investigation. The galaxy mock catalogue is extracted 
from the $N$-body simulation (\cf~\S~\ref{subsection:makingmocks}). 

\subsection{Making Mock Catalogues}\label{subsection:makingmocks}
Mock \pscz\, catalogues have been extracted from the $N$-body
simulations performed by Cole \etal~(1997).  They used the AP$^3$M 
code of Couchman (1991) loaded with 192$^3$ particles in a box of 
comoving size of 345.6$\hm$. The particle mass is 1.62$\times$10$^{12}\,\Omega_{M} h^{-1} M_{\odot}$. 
Further details can be found in Cole \etal~(1997). For the analysis performed in 
\S~\ref{applications} we have considered two different cosmologies 
(\cf~Table \ref{t:models1}): a flat model $\Omega_{M}=0.3$ and 
cosmological constant term, $\Omega_{\Lambda} c^2/3H_0^2=0.7$ and a critical 
density universe ($\Omega_{M} = 1.0$) with power spectrum shape parameter, 
$\Gamma=0.25$. The relevant details of the two cosmological models explored are 
summarised in the table below.
\begin{table}[h]
\begin{center}
\label{t:models1}
\tabcolsep 1pt
\begin{tabular}{lcccc}  \\ \hline \hline
 \\
~~Model~~~ &~~~$\Omega_{M}$~~ & ~~~$\Omega_{\Lambda}$~~ &~~~$\Gamma$~~ & ~~~$\sigma_8$~~ \\ 
\hline
\\
~~LCDM & 0.3 & 0.7 & 0.25 & 1.13 \\ 
~~SCDMG & 1.0 & 0.0 & 0.25 & 0.55 \\ 
 \\
\hline \hline
\end{tabular}
\caption[Cosmological Models]
	{\small Cosmological Models.}
\end{center}
\end{table}

Ten different mock catalogues, which we will refer to as LCDM0$i$,
SCDMG0$i$, $i=0,...,9$, have been extracted from each of the above models.  

Although in this work we have not created an $N$-body code or extracted the mock 
catalogues from it, it is still worth discussing the procedure for doing so.
\begin{itemize} 
\item{ }A population of particles with properties similar to those of
  the Local Group (LG) is identified.  A LG-like observer is defined by
  implementing two observationally based constraints: the peculiar
  velocity of the point must be ${\vvec}_{LG}= 625 \pm 25$\kms, and the
  particle must be located in a region for which the fractional
  overdensity $\delta$ averaged in a radius of 5$\hm$ is in the range
  $-0.2 < \delta < 1.0 $.
\item{ } A sphere of 120$\hm$ radius is drawn around the LG-like
  observer and the whole frame is rotated so that the motion of the
  observer is in the direction ($l=276^{\circ}$,\,$b=30^{\circ}$), the
  direction of the LG peculiar velocity with respect to the CMB frame
  (e.g. Wilkinson 1988).
\item{ } A friends-of-friends algorithm is implemented to find
  galaxy clusters. (see Frenk \etal~1988 for details)
\item{ }The number density of particles in the simulation is $\sim
  0.039$~$h^{3}$Mpc$^{-3}$~while the number density given by the
  \pscz~selection function exceeds this density closer than some
  critical distance.  Thus, the simulations are volume-limited for
  distances less than 10.9$\hm$, where the \pscz~number density
  (Saunders \etal~1999) matches the $N$-body one. For
  distances greater than this the simulated surveys follow the \pscz\,
  number density.
\item{ }A Monte Carlo rejection was used to choose particles according
  to the \pscz~selection function (Saunders
  \etal~2000):
\begin{equation}
\label{eq:sel_fun3}
\phi(r) = 
\left( {r_{\circ}}\over{r} \right)^{2\alpha}
\left (
{r_{\star}^2 + r_{\circ}^2}
\over
{ r_{\star}^2+ {r}^2} 
\right) ^{\beta}.
\end{equation}
The optimal parameters are listed in Table~\ref{t:self}.  For $r \le
10.9 \hm$ the mock catalogues are volume-limited and thus $\phi(r)=1$.
A random flux consistent with the \pscz~selection function is then
attributed at each selected galaxy.
\begin{table}[h]
\begin{center}
\tabcolsep 1pt
\begin{tabular}{cccc} \hline \hline \\
  {~~~$\alpha$~~~}  & {~~~$\beta$~~~}  &~~~~$r_{\circ}$~~~& ~~~$r_{\star}$~~~ \\  
                    &                  &        ~$[\hm]$~ &         ~$[\hm]$~ \\
\hline
0.53 & 1.90 & 10.90 & 86.40 \\
\hline \hline
\end{tabular}
\label{t:self}
\caption{\small Selection Function Parameters for Eq.~(\ref{eq:sel_fun3})}
\end{center}
\end{table}
\item{ }
Despite the large sky coverage, \pscz\, is not a full-sky catalogue.
Unsurveyed regions are present both at high and low  galactic
latitudes that need to be accounted for to properly reproduce the
existing observational biases. All the galaxies which fall within masked 
regions have been rejected leaving the final sky coverage to be $\approx 84 \%$
complete.
\end{itemize}


The final mock catalogue contains the positions of the galaxies in 
redshift-space and their ``observed'' flux.  The galaxy redshifts are 
assigned by adding the line-of-sight component of the peculiar velocity 
to the recession velocity. 

\subsection{Multifractal Analysis}
Figure \ref{fig:3d-fit}~shows the $D_{q,R}$ surface for the halo positions in 
an ideal and complete $(400~h^{-1}\rm{Mpc})^{3}$ box. It has a standard $\Lambda$CDM 
cosmology with no galaxies, so only the underlying dark matter distribution 
was investigated. Overall, it is a very smooth surface, which tends 
towards homogeneity at scales $> 30 h^{-1}$Mpc. A clear peak, at low $q$, and dip, 
at high $q$ corresponds to multifractality for $R \approx 10 h^{-1}$Mpc.

\begin{figure}
\begin{center}
\includegraphics[height=9cm,width=16cm]{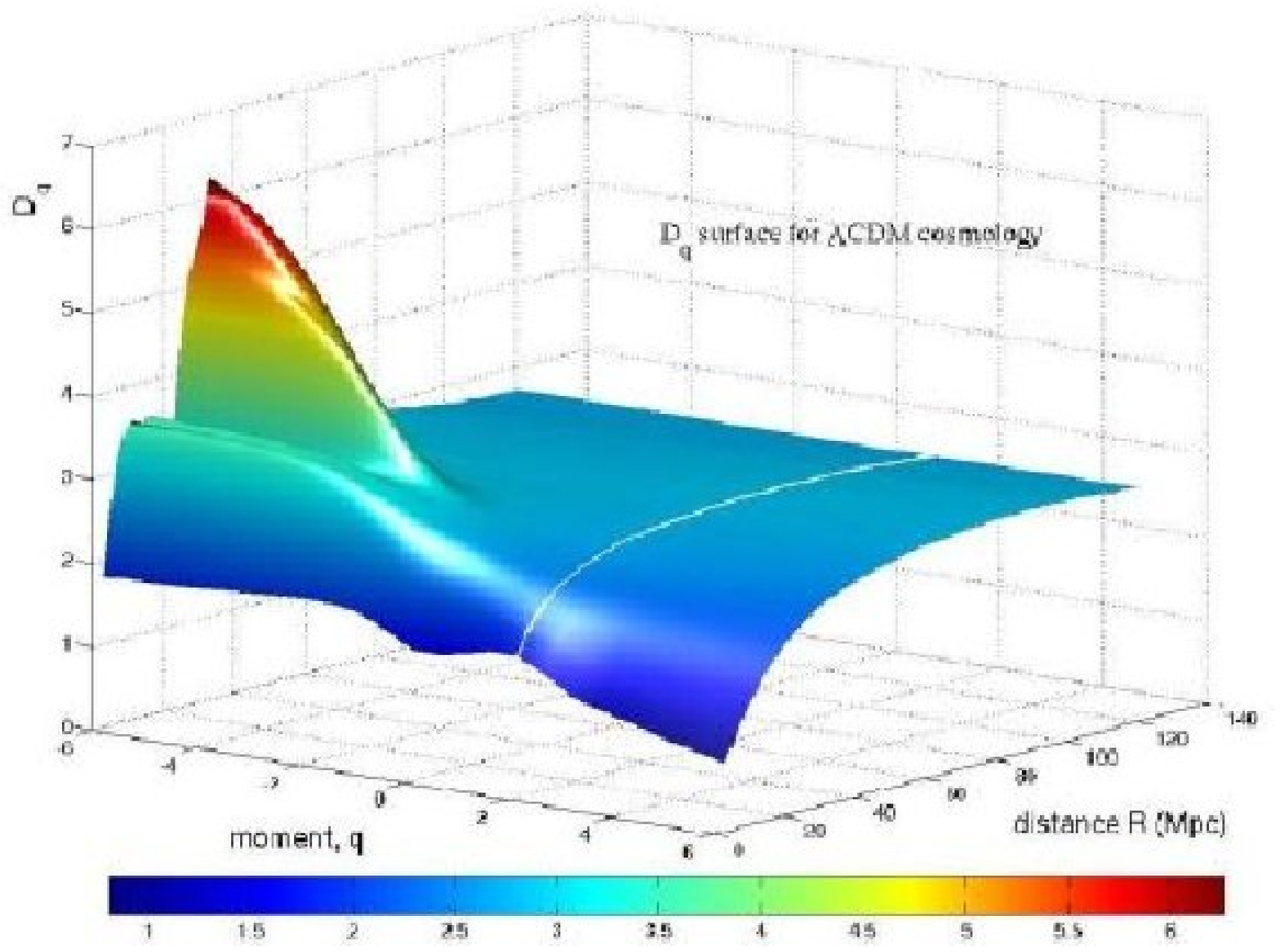}\\
\includegraphics[height=9cm,width=15cm]{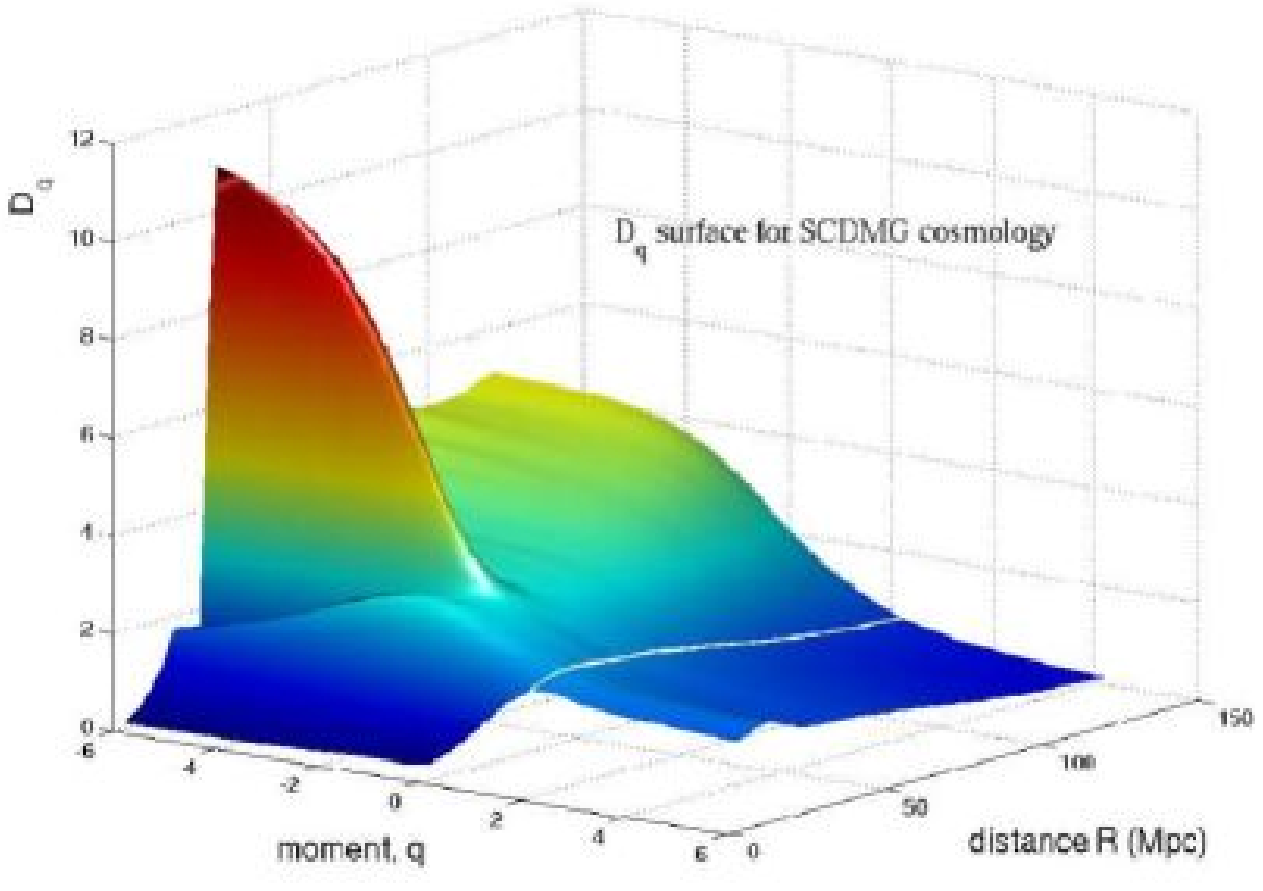}
\caption[$D_{q,R}$ surface plot comparing two simulations: $\Lambda$CDM \& SCDM]
	{\small $D_{q,R}$ surfaces for two different cosmologies. The white reference line corresponds 
to the correlation dimension, $D_{2}$, which is usually obtained from the derivative of the 2PCF. 
{\em Top}: A $\Lambda$CDM cosmology. {\em Bottom}: An SCDMG cosmology. The two white reference 
curves are very similar on small scales in that they give the usual value of $D_{2} \sim 2$. On 
larger scales, however, it is evident that the $D_{q,R}$ surface can in principle differentiate 
between different cosmological models.}
\label{fig:two_cosmologies}
\end{center}
\end{figure}

The same $D_{q,R}$ surface analysis is repeated, this time for the Mock galaxy 
catalogues mentioned above. The main difference is, now there are a lot less 
particles to analyse, $\sim 15,000$. However, given that we have 10 mock 
realisations to average over, the noise should not be much worse. Figure 
\ref{fig:two_cosmologies} shows the averaged $D_{q,R}$ surfaces of the $\Lambda$CDM 
and the SCDMG cosmologies. 

\subsection*{$D_{q,R}$ Surface - $\Lambda$CDM}
There are some points we should note regarding the $D_{q,R}$ surfaces of figure 
\ref{fig:two_cosmologies}. The $\Lambda$CDM cosmology yields a plot quite similar 
to that of figure \ref{fig:3d-fit}, this should not be too surprising given that 
they have the same cosmological parameters. There is however clearly a few 
features on small scale where the two differ. In fig.(\ref{fig:two_cosmologies}) 
at $\sim 10 h^{-1}$Mpc there is a multifractal feature peaking at $D_{q=-6}\approx3.5$. 
A little further out at $\sim 20 h^{-1}$Mpc there is a second multifractal signature with 
a much higher peak $D_{q=-6}\approx6.5$. Recalling that figure 
\ref{fig:two_cosmologies} is an average taken over 10 realisations, it is interesting 
that the two multifractal features on small scales have not merged into one \ie~they 
are probably not statistical anomalies. It would be premature to suggest that these 
distinct features have any physical significance\footnote{This same point has been 
made by Bernard Jones in many of his papers, most recently in the review article 
(Jones \etal~,2005).} but there is definitely room here for further study.

On scales larger than $30 h^{-1}$Mpc the surface flattens to homogeneity, $D_{q}\rightarrow~3$. 
The white reference line shows the correlation dimension, $D_{2}$, varying with scale. 
This is related to the derivative of the 2PCF. Visually this shows that all the 
information which the 2PCF can provide is but a very small part of the $D_{q,R}$ 
surface.

\subsection*{$D_{q,R}$ Surface - SCDMG}
In the lower panel of figure \ref{fig:two_cosmologies}, the $D_{q,R}$ surface is 
plotted for the SCDMG cosmology. It clearly shows on small scales a similar multifractal 
peak to that of the $\Lambda$CDM model. However, there seems to be only one very 
sharp distinct peak at $\sim 20 h^{-1}$Mpc and for $q=6$, $D_{q=6}\approx 11$. Casting 
our mind back to \S\ref{MFA}, this is telling us that on scales $\sim 20 h^{-1}$Mpc void 
under-dense regions are more clustered.

Looking beyond the first peak it can be seen that the surface does not flatten 
off like the $\Lambda$CDM model. In fact on scales beyond $30 h^{-1}$Mpc there is no 
transition to homogeneity in this universe, on scales up to the size of the simulation 
it is entirely multifractal.

It is evident, therefore, that on large scale the $D_{q,R}$ surface can in principle 
discriminate between different cosmological models.

\chapter{Summary}
\vspace{-0.6cm}
We have shown that our {\em volume} correction recovers the true fractal dimension, 
when tested with the extreme case of an anisotropic L\'{e}vy Flight. The increase in 
accuracy over Pan \& Coles angular correction estimator is marginal, however the 
computational load required is much less. This fact will become increasingly more 
important as red-shift surveys are now pushing 1,000,000+ galaxies.

We present the $D_{q,R}$ surface as a possible unique descriptor of a discrete point 
distribution. Whether this is strictly true or not does not differ from the fact 
that there is much more information contained in our multifractal analysis than can 
be extracted from the usual 2PCF. In fact to go from a 2PCF to our multifractal 
measure, is of no significant computational cost.
 
The 2PCF in our formalism, is represented by the integral along the $q=2$ line in 
figures (\ref{fig:two_cosmologies}) \& (\ref{fig:3d-fit}). It is clearly apparent 
that $q=2$ is confined to a rather boring and flat part of the surface and generally 
for $q > 1$ the surface is very smooth and $D_{q}$ tends towards a constant value 
without any interesting features.

There has been a lot of interest, since the dawn of cosmological simulations, to 
apply statistics to the resulting distributions so that a comparison can be made 
with real surveys. Much of this effort has involved the 2PCF (also Minimal Spanning 
Trees and other geometrically motivated descriptors). However it must be made clear 
that the trade off between statistical robustness and visual interpretation is of 
prime importance when quantifying structure. Neither should be favoured too heavily.

As we have presented here, our $D_{q,R}$ surface gives varying weights to under dense 
(void) regions and over dense (clustered) regions, this obviously has a significant 
advantage over calculating higher order correlations. The major task now, regarding 
the $D_{q,R}$ surface, is to make full use of it and to extracted as much information 
as is statistically possible.

Our methodology as presented in \S\ref{MFA} also has the possibility to contribute 
to parameter estimation. Especially in the area of Baryon Acoustic Oscillations (BAO). 
BAO's are difficult to measure accurately due to the low amplitude on scale above 
100Mpc. To help increase the power in this range, clusters, instead of individual 
galaxies, could be used as they show a higher clustering amplitude. This technique 
has some drawbacks The biggest problem it faces is in accurately determining what is 
a cluster. Another technique for measuring BAO's would be to use the $D_{q,R}$ surface 
as we introduced in \S\ref{LCDM_MFA}. Since this method treats dense and under-dense 
regions differently, it suggests that the usual 2PCF (or $Z(2,r)$) may not be the best 
suited to observing and measuring BAO's. In future work we hope to explore the use of 
the $D_{q,R}$ surface as a cosmological tool.
\end{spacing}

\nocite{*}

\begin{thebibliography}{10}

\bibitem{Aarseth:1978}
S.~J. {Aarseth}.
\newblock {Computer simulations of galaxy clustering}.
\newblock In M.~S. {Longair} and J.~{Einasto}, editors, {\em IAU Symp. 79:
  Large Scale Structures in the Universe}, pages 189--194, 1978.

\bibitem{abazajian+:2005}
K.~{Abazajian \etal }.
\newblock {The Third Data Release of the Sloan Digital Sky Survey}.
\newblock {\em \aj}, 129, March 2005.

\bibitem{abell:1958}
G.~O. {Abell}.
\newblock {The Distribution of Rich Clusters of Galaxies.}
\newblock {\em \apjs}, 3(211), May 1958.

\bibitem{barrow+:1985}
J.~D. {Barrow}, S.~P. {Bhavsar}, and D.~H. {Sonoda}.
\newblock {Minimal spanning trees, filaments and galaxy clustering}.
\newblock {\em \mnras}, 216:17--35, September 1985.

\bibitem{bertschinger+:1991}
E.~{Bertschinger} and J.~M. {Gelb}.
\newblock {Cosmological N-body simulations}.
\newblock {\em Computers in Physics}, 5:164--175, April 1991.

\bibitem{Boldyrev+:2003}
S.~{Boldyrev} and C.~R. {Gwinn}.
\newblock {L{\'e}vy Model for Interstellar Scintillations}.
\newblock {\em Physical Review Letters}, 91(13):131101, September 2003.

\bibitem{Borgani+:1994}
S.~{Borgani}, V.~J. {Martinez}, M.~A. {Perez}, and R.~{Valdarnini}.
\newblock {Is there any scaling in the cluster distribution?}
\newblock {\em \apj}, 435, November 1994.

\bibitem{Carroll:book}
S.~M. {Carroll}.
\newblock {\em {Spacetime and geometry. An introduction to general
  relativity}}.
\newblock Spacetime and geometry / San Francisco, CA, USA: Addison Wesley, ISBN
  0-8053-8732-3, 2004, 2004.

\bibitem{Chowdhury+:1999}
D.~Chowdhury and D.~Stauffer.
\newblock A generalized spin model of financial markets.
\newblock {\em Euro. Phys Journal B}, 8:477, 1999.

\bibitem{Clark+:1966}
R.~{Clark} and W.F. {Miller}.
\newblock {\em Math. Comp. Phy}, 1966.

\bibitem{cole+:1997}
S.~{Cole}, D.~H. {Weinberg}, C.~S. {Frenk}, and B.~{Ratra}.
\newblock {Large-scale structure in COBE-normalized cold dark matter
  cosmogonies}.
\newblock {\em \mnras}, 289, July 1997.

\bibitem{colless:+1999}
M.~{Colless}.
\newblock {First results from the 2dF Galaxy Redshift Survey}.
\newblock In G.~{Efstathiou}, editor, {\em Large-Scale Structure in the
  Universe}, 1999.

\bibitem{couchman:1991}
H.~M.~P. {Couchman}.
\newblock {Mesh-refined P3M - A fast adaptive N-body algorithm}.
\newblock {\em \apjl}, 368, February 1991.

\bibitem{davis-peebles:1983}
M.~{Davis} and P.~J.~E. {Peebles}.
\newblock {A survey of galaxy redshifts. V - The two-point position and
  velocity correlations}.
\newblock {\em \apj}, 267:465--482, April 1983.

\bibitem{Doroshkevich+:2004}
A.~{Doroshkevich}, D.~L. {Tucker}, S.~{Allam}, and M.~J. {Way}.
\newblock {Large scale structure in the SDSS galaxy survey}.
\newblock {\em \aap}, 418:7--23, April 2004.

\bibitem{efstathiou+:1985}
G.~{Efstathiou}, M.~{Davis}, S.~D.~M. {White}, and C.~S. {Frenk}.
\newblock {Numerical techniques for large cosmological N-body simulations}.
\newblock {\em \apjs}, 57:241--260, February 1985.

\bibitem{frenk+:1988}
C.~S. {Frenk}, S.~D.~M. {White}, M.~{Davis}, and G.~{Efstathiou}.
\newblock {The formation of dark halos in a universe dominated by cold dark
  matter}.
\newblock {\em \apj}, 327, April 1988.

\bibitem{gott+:1979}
J.~R. {Gott}, III and E.~L. {Turner}.
\newblock {An extension of the galaxy covariance function to small scales}.
\newblock {\em \apjl}, 232:L79--L81, September 1979.

\bibitem{grassberger+:1983}
P.~{Grassberger} and I.~{Procaccia}.
\newblock {Estimation of the Kolmogorov entropy from a chaotic signal}.
\newblock {\em \pra}, 28, October 1983.

\bibitem{Hatton:1999}
S.~{Hatton}.
\newblock {Approaching a homogeneous galaxy distribution: results from the
  Stromlo-APM redshift survey}.
\newblock {\em \mnras}, 310:1128--1136, December 1999.

\bibitem{heath:1977}
D.~J. {Heath}.
\newblock {The growth of density perturbations in zero pressure
  Friedmann-Lemaitre universes}.
\newblock {\em \mnras}, 179:351--358, May 1977.

\bibitem{hentschel+:1983}
H.~G.~E. {Hentschel} and I.~{Procaccia}.
\newblock {Fractal nature of turbulence as manifested in turbulent diffusion}.
\newblock {\em \pra}, February 1983.

\bibitem{hinshaw+:2003}
G.~{Hinshaw}, D.~N. {Spergel}, L.~{Verde}, R.~S. {Hill}, S.~S. {Meyer},
  C.~{Barnes}, C.~L. {Bennett}, M.~{Halpern}, N.~{Jarosik}, A.~{Kogut},
  E.~{Komatsu}, M.~{Limon}, L.~{Page}, G.~S. {Tucker}, J.~L. {Weiland},
  E.~{Wollack}, and E.~L. {Wright}.
\newblock {First-Year Wilkinson Microwave Anisotropy Probe (WMAP) Observations:
  The Angular Power Spectrum}.
\newblock {\em \apjs}, 148:135--159, September 2003.

\bibitem{huchra+:1983}
J.~{Huchra}, M.~{Davis}, D.~{Latham}, and J.~{Tonry}.
\newblock {A survey of galaxy redshifts. IV - The data}.
\newblock {\em \apjs}, 52, June 1983.

\bibitem{jones+:2005}
B.~J. {Jones}, V.~J. {Mart{\'{\i}}nez}, E.~{Saar}, and V.~{Trimble}.
\newblock {Scaling laws in the distribution of galaxies}.
\newblock {\em Reviews of Modern Physics}, 76:1211--1266, February 2005.

\bibitem{Jones:1988}
B.~J.~T. {Jones}, V.~J. {Martinez}, E.~{Saar}, and J.~{Einasto}.
\newblock {Multifractal description of the large-scale structure of the
  universe}.
\newblock {\em \apjl}, 332, September 1988.

\bibitem{Kerscher+:2000}
M.~{Kerscher}, I.~{Szapudi}, and A.~S. {Szalay}.
\newblock {A Comparison of Estimators for the Two-Point Correlation Function}.
\newblock {\em \apjl}, 535:L13--L16, May 2000.

\bibitem{komatsu+:2003}
E.~{Komatsu}, A.~{Kogut}, M.~R. {Nolta}, C.~L. {Bennett}, M.~{Halpern},
  G.~{Hinshaw}, N.~{Jarosik}, M.~{Limon}, S.~S. {Meyer}, L.~{Page}, D.~N.
  {Spergel}, G.~S. {Tucker}, L.~{Verde}, E.~{Wollack}, and E.~L. {Wright}.
\newblock {First-Year Wilkinson Microwave Anisotropy Probe (WMAP) Observations:
  Tests of Gaussianity}.
\newblock {\em \apjs}, 148, September 2003.

\bibitem{Landy+:1993}
S.~D. {Landy} and A.~S. {Szalay}.
\newblock {Bias and variance of angular correlation functions}.
\newblock {\em \apj}, 412:64--71, July 1993.

\bibitem{Loveday+:1992}
J.~{Loveday}, B.~A. {Peterson}, G.~{Efstathiou}, and S.~J. {Maddox}.
\newblock {The Stromlo-APM Redshift Survey. I - The luminosity function and
  space density of galaxies}.
\newblock {\em \apj}, 390, May 1992.

\bibitem{Ma+:1995}
C.-P. {Ma} and E.~{Bertschinger}.
\newblock {Cosmological Perturbation Theory in the Synchronous and Conformal
  Newtonian Gauges}.
\newblock {\em \apj}, 455, December 1995.

\bibitem{maddox+:1990}
S.~J. {Maddox}, G.~{Efstathiou}, W.~J. {Sutherland}, and J.~{Loveday}.
\newblock {The APM galaxy survey. I - APM measurements and star-galaxy
  separation}.
\newblock {\em \mnras}, 243:692--712, April 1990.

\bibitem{mandelbrot:book}
B.~B. {Mandelbrot}.
\newblock {\em {The Fractal Geometry of Nature}}.
\newblock The Fractal Geometry of Nature, San Francisco: Freeman, 1982, 1982.

\bibitem{martinez+:1990}
V.~J. {Martinez}, B.~J.~T. {Jones}, R.~{Dominguez-Tenreiro}, and R.~{van de
  Weygaert}.
\newblock {Clustering paradigms and multifractal measures}.
\newblock {\em \apj}, 357:50--61, July 1990.

\bibitem{Meakin:book}
P.~{Meakin}.
\newblock {\em {Fractal, scaling and growth far from equilibrium}}.
\newblock Cambridge University Press, 1998.

\bibitem{meiksin+:1992}
A.~{Meiksin}, I.~{Szapudi}, and A.~{Szalay}.
\newblock {Higher order correlations of IRAS galaxies}.
\newblock {\em \apj}, 394:87--90, July 1992.

\bibitem{gravitation:1973}
C.~W. {Misner}, K.~S. {Thorne}, and J.~A. {Wheeler}.
\newblock {\em {Gravitation}}.
\newblock San Francisco: W.H.~Freeman and Co., 1973, 1973.

\bibitem{PC:2000}
J.~{Pan} and P.~{Coles}.
\newblock {Large-scale cosmic homogeneity from a multifractal analysis of the
  PSCz catalogue}.
\newblock {\em \mnras}, 318:L51--L54, November 2000.

\bibitem{PC:2002}
J.~{Pan} and P.~{Coles}.
\newblock {Boundary corrections in fractal analysis of galaxy surveys}.
\newblock {\em \MNRAS}, March 2002.

\bibitem{peebles:book:1980}
P.~J.~E. {Peebles}.
\newblock {\em {The large-scale structure of the universe}}.
\newblock Princeton University Press, 1980.~435 p., 1980.

\bibitem{peebles:book:1993}
P.~J.~E. {Peebles}.
\newblock {\em {Principles of physical cosmology}}.
\newblock Princeton University Press, 1993.

\bibitem{percival+:2006}
W.~J. {Percival \etal~}.
\newblock {The shape of the SDSS DR5 galaxy power spectrum}.
\newblock {\em ArXiv Astrophysics e-prints}, August 2006.

\bibitem{Saslaw:book}
W.~C. {Saslaw}.
\newblock {\em {The Distribution of the Galaxies}}.
\newblock The Distribution of the Galaxies.~ISBN 0521394260. Cambridge
  University Press, October 1999.

\bibitem{saunders+:2000}
W.~{Saunders}, S.~{Oliver}, O.~{Keeble}, M.~{Rowan-Robinson}, S.~{Maddox},
  R.~{McMahon}, G.~{Efstathiou}, W.~{Sutherland}, H.~{Tadros}, S.~D.~M.
  {White}, and C.~S. {Frenk}.
\newblock {The IRAS Point Source Catalog Redshift (PSCz) Survey}.
\newblock In {\em ASP Conf. Ser.}, 2000.

\bibitem{seljak+:1996}
U.~{Seljak} and M.~{Zaldarriaga}.
\newblock {A Line-of-Sight Integration Approach to Cosmic Microwave Background
  Anisotropies}.
\newblock {\em \apj}, 469, October 1996.

\bibitem{springel+:2005}
V.~{Springel}, S.~D.~M. {White}, A.~{Jenkins}, C.~S. {Frenk}, N.~{Yoshida},
  L.~{Gao}, J.~{Navarro}, R.~{Thacker}, D.~{Croton}, J.~{Helly}, J.~A.
  {Peacock}, S.~{Cole}, P.~{Thomas}, H.~{Couchman}, A.~{Evrard}, J.~{Colberg},
  and F.~{Pearce}.
\newblock {Simulations of the formation, evolution and clustering of galaxies
  and quasars}.
\newblock {\em nature}, 435, June 2005.

\bibitem{strauss+:1995}
M.~A. {Strauss} and J.~A. {Willick}.
\newblock {The density and peculiar velocity fields of nearby galaxies}.
\newblock {\em \physrep}, 261:271--431, 1995.

\bibitem{venables+:1999}
W.~N. Venables and B.~D. Ripley.
\newblock {\em Modern Applied Statistics with S-PLUS}.
\newblock Springer, New York, 1999.

\bibitem{Virey+:2005}
J.-M. {Virey}, P.~{Taxil}, A.~{Tilquin}, A.~{Ealet}, C.~{Tao}, and
  D.~{Fouchez}.
\newblock {Determination of the deceleration parameter from supernovae data}.
\newblock {\em \prd}, 72(6):061302, September 2005.

\bibitem{warren+:2006}
M.~S. {Warren}, K.~{Abazajian}, D.~E. {Holz}, and L.~{Teodoro}.
\newblock {Precision Determination of the Mass Function of Dark Matter Halos}.
\newblock {\em \apj}, 646, August 2006.

\bibitem{weinberg:book}
S.~{Weinberg}.
\newblock {\em {Gravitation and Cosmology: Principles and Applications of the
  General Theory of Relativity}}.
\newblock ISBN 0-471-92567-5.~Wiley-VCH, July 1972.

\bibitem{wilkinson:1988}
D.~T. {Wilkinson}.
\newblock {Recent Measurements of the Cosmic Microwave Radiation}.
\newblock In J.~{Audouze}, M.-C. {Pelletan}, and S.~{Szalay}, editors, {\em IAU
  Symp. 130: Large Scale Structures of the Universe}, 1988.

\bibitem{zwicky:1968}
F.~{Zwicky \etal }.
\newblock {Catalogue of Galaxies and of Clusters of Galaxies. vol. 6.}
\newblock In {\em California Inst. Techn.}, 1968.

\end{thebibliography}

\end{document}